\documentclass[12pt,preprint]{aastex}

\newcommand{\brbp}{\langle B_R^2\rangle/\langle B_\phi^2\rangle}
\newcommand{\Qz}{\langle Q_z \rangle}
\newcommand{\Qp}{\langle Q_\phi \rangle}

\begin{document} 
\title{Testing Convergence for Global Accretion Disks}

\author{John F. Hawley, Sherwood A. Richers, Xiaoyue Guan}
\affil{Department of Astronomy \\ University of Virginia \\ P.O. Box
400325 \\
Charlottesville, VA 22904-4325}
\email{jh8h@virginia.edu; xg3z@virginia.edu}

\and
\author{Julian H. Krolik}
\affil{Department of Physics and Astronomy\\
Johns Hopkins University\\
Baltimore, MD 21218}
\email{jhk@pha.jhu.edu}

\begin{abstract}

Global disk simulations provide a powerful tool for investigating
accretion and the underlying magnetohydrodynamic turbulence driven by
magneto-rotational instability (MRI).  Using them to predict accurately
quantities such as stress, accretion rate, and surface brightness profile
requires that purely numerical effects, arising from both resolution and
algorithm, be understood and controlled.  We use the flux-conservative
{\it Athena} code to conduct a series of experiments on disks having a
variety of magnetic topologies to determine what constitutes adequate
resolution.  We develop and apply several resolution metrics:  $\Qz$
and $\Qp$, the ratio of the grid zone size to the characteristic MRI
wavelength, $\alpha_{mag}$, the ratio of the Maxwell stress to the
magnetic pressure, and $\brbp$, the ratio of radial to toroidal magnetic
field energy.  For the initial conditions considered here, adequate
resolution is characterized by $\Qz \ge 15$, $\Qp \ge 20$, $\alpha_{mag}
\approx 0.45$, and $\brbp \approx 0.2$.  These values are associated with
$\ge 35$ zones per scaleheight $H$, a result consistent with shearing
box simulations.  Numerical algorithm is also important.  Use of the HLLE
flux solver or second-order interpolation can significantly degrade the
effective resolution compared to the HLLD flux solver and third-order
interpolation.  Resolution at this standard can be achieved only with
large numbers of grid zones, arranged in a fashion that matches the
symmetries of the problem and the scientific goals of the simulation.
Without it, however, quantitative measures important to predictions of
observables are subject to large systematic errors.

\end{abstract}


\section{Introduction}

Accretion is one of the fundamental power sources in astrophysical
systems.  Although the basic theory of accretion disks was formulated
decades ago, it is not yet possible to carry out detailed modeling
of accretion systems from first principles.  The physics involved
is too complex, and the problem is inherently multi-dimensional and
time-dependent.  The rapid increase in high performance computing
capabilities has made possible increasingly detailed three-dimensional
simulations, both in Newtonian (or pseudo-Newtonian) gravity and in
the fully relativistic space-time of a Kerr black hole.  The results
of those simulations, however, require careful interpretation if they
are to be usefully applied.  The outcome of a given simulation
depends not only on one's choices in physical properties, such as black
hole spin and mass, and the density, temperature, and angular momentum
of the accreting gas, but also on model-dependent factors such as the
initial and boundary conditions, computational domain size, numerical
resolution, and numerical algorithm.  Of all these factors, those that
are purely numerical can be the most vexing as their influences are both
unintended and uncontrolled.  One can be somewhat less concerned with
the influence of numerical factors while conducting an initial {\it
qualitative} survey of possible accretion behaviors and properties.
But as we begin to answer more detailed {\it quantitative} questions
that have the potential to relate more closely to observations, it
becomes increasingly necessary to develop procedures to evaluate
the merits of different simulations and the degree to which the numbers
they provide can be regarded as accurate.

The accuracy of a given simulation is often described in terms of the {\it
convergence} of the simulation.  The terminology is imprecise, at least as
it is often applied to accretion simulations.  True numerical convergence
means that the truncation errors in the solution are approaching zero at
the rate consistent with the order of the scheme and that there are no
remaining unresolved structures at the grid zone scale.  In that case,
a simulation result does not change as resolution is further increased.
Formal convergence cannot be achieved for ideal magnetohydrodynanic (MHD),
or hydrodynamic, simulations of turbulence that lack fixed viscous and resistive
lengthscales; there will always be new structure created as resolution
is increased.  Instead, what is usually meant by convergence in the
context of such simulations is that physically important, macroscopic
quantitative values (such as accretion rate and stress) do not change by
a significant amount as the numerical resolution is increased \citep[see
also the discussion in][]{Sorathia:2012}

Even this less formal standard of convergence is hard to demonstrate
or achieve in three-dimensional time-dependent simulations.  First,
it is generally impractical to employ ever-increasing resolution;
one typically carries out simulations with as high a resolution as
one can.  Lower resolution simulations might provide useful data,
but the interpretation may be ambiguous.  If the best resolution one
can achieve is still under-resolved, then the changes observed in lower
resolution simulations may not be sufficient to determine how much more
a simulation would change at ever higher resolution.  The absence of a
significant difference between a high and a lower resolution simulation
is also not conclusive.  If a simulation on a fine grid does not resolve
an important process, neither will simulations done on a coarser grid.

Given these somewhat daunting practical difficulties, how can
we judge the adequacy of a global accretion disk simulation?
In accretion, the key dynamic quantity is the magnitude of the
$r$--$\phi$ stress $\tau_{r\phi}$ that transports angular momentum
outward through the disk.  This is usually characterized in terms
of the Shakura-Sunyaev \citep{Shakura:1973} parameter $\alpha
= \tau_{r\phi}/P$, where $P$ is the thermal pressure.  Both the
Maxwell and Reynolds stresses that make up $\tau_{r\phi}$ originate
in MHD turbulence driven by the magneto-rotational instability
\citep[MRI;][]{Balbus:1991,Balbus:1992,Balbus:1998}.  Turbulence is
notoriously difficult to compute adequately even in a local domain;
the global problem is much more challenging.  For accretion, the local
domain is the ``shearing box'' \citep{Hawley:1995b} which uses Cartesian
geometry and includes the local tidal and Coriolis forces appropriate to
differential rotation.  There are two widely used versions of shearing
box:  stratified and unstratified, depending on whether or not they
include the vertical component of gravity.
The shearing box is characterized by an angular
frequency $\Omega$ and the isothermal pressure scale height $H$.
Different groups define $H$ differently. Where it is necessary 
to distinguish usage, we label $H$ as follows:
for an isothermal sound speed  $c_s$, $H_1=c_s/\Omega$ and $H_2=\sqrt{2}
c_s/\Omega$.  For the stratified shearing box $H$ is the vertical
scale height; in unstratified simulations it simply sets a
characteristic scale based on the sound speed.
Other important shearing box parameters include
the box domain size, typically a few $H$ in each dimension,
and the initial magnetic
field topology and strength, measured by $\beta = P_{gas}/P_{mag}$.
Here again there is variation in how $\beta$ is defined, e.g. in terms
of average values or values at the equatorial plane.

The local nature of the shearing box makes it possible to use a large number of
grid zones per $H$. Some recent shearing box models used 
32--128 zones per $H_2$ in \cite{Davis:2010},  32--144 zones
per $H_2$ in \cite{Simon:2012}, and 12.8 zones per $H_1$ in
\cite{Guan:2011}.
These local simulations find $\alpha
\sim 0.01$--$0.1$; \cite{Simon:2012},
found $\alpha \approx 0.02$--$0.03$  to be typical,
with considerable temporal and spatial variation, while the less
well-resolved simulations of \cite{Guan:2011} report $\alpha =
0.01$--0.02.  Work continues in the local
box models to better understand resolution effects, as well as
the impacts of non-zero resistivity and
viscosity \citep{Fromang:2007b,Lesur:2007,Simon:2009b}.

It is hoped that the insights gained from studies of MRI-driven turbulence
in local models will carry over to the global domain.  To establish that
this is so, we need first to determine whether the local shearing box is
representative of the local behavior within a global disk.  Since the
MRI is a local instability, it seems reasonable to assume that the
properties of turbulence observed in the shearing box are applicable in
global contexts.  The recent work of \cite{Simon:2012} on large shearing
boxes suggests that this is the case for some properties of the turbulence
(e.g. $\alpha$), but that there are also what they refer to as
``mesoscale'' structures on scales $\sim 10H$.  On the other hand, when
shearing-boxes are sufficiently large, the curvature of real disks may
become important.

Next, what resolutions do global simulations require, and can shearing box
results be used to determine this?  Because accretion disks span large
radial distances and are thin, $H/R < 1$, it is extremely difficult
for global models to achieve anything close to the resolution used
in local simulations.  The approach taken by \cite{Hawley:2011} was to
characterize the resolution-dependence of certain average properties of
MRI-driven turbulence in well-resolved local shearing box simulations.
Characteristic properties that approach a fixed value as resolution
increases then become diagnostics of quality and can be used as a standard
against which global simulations can be evaluated.  Based on comparisons
to well resolved shearing box simulations, \cite{Hawley:2011} concluded
that no global simulations have yet been run with sufficient resolution
as to be deemed ``converged'' in certain essential MHD properties.

The first of these diagnostics derives from the characteristic wavelength
of the MRI mode, $\lambda_{MRI} = 2\pi v_a/\Omega$, where $v_a$ is the
Alfv\'en speed; $\lambda_{MRI}$ is the distance an Alfv\'en wave travels in one orbit.  This definition differs from the precise fastest
growing wavelength by a factor close to unity.  From this definition
\cite{Hawley:2011} define two simulation quality metrics from the ratio
of this length to the grid zone size, namely
\begin{equation}\label{eq:qzdef}
Q_z = {\lambda_{MRI} \over \Delta z} = \frac{2 \pi |v_{az}|}{\Omega \Delta z} ,
\end{equation}
where $v_{az}$ is the $z$ component of the Alfv\'en speed, and
\begin{equation}\label{eq:qydef}
 Q_\phi = {\lambda_{MRI}\over R\Delta\phi } = \frac{2\pi
|v_{a\phi}|}{R \Omega \Delta \phi}
= {2\pi
H_2\over(\beta_\phi^{1/2} R\Delta \phi)}, 
\end{equation}
which usefully defines $Q_\phi$ in terms of the local scale height $H$
and $\beta_\phi$, the ratio of the gas to magnetic pressure including only
the contribution from the toroidal field.  The importance of resolving
the characteristic MRI wavelength has long been recognized.  Previous
numerical tests \citep[e.g.,][]{Hawley:1995,Sano:2004,Fromang:2006}
indicate that obtaining proper {\it linear} growth rates for
the fastest-growing MRI modes requires $> 6$ zones per wavelength.
\cite{Flock:2010} carried out a study of the linear MRI
in a global disk configuration and concluded that $Q_z > 8$ was
required to accurately capture the growth rates, at least for
numerical schemes that were relatively low in numerical diffusion.
Although the $Q$ parameters were invented to calibrate the description
of linear amplitude MRI, they continue to be useful diagnostics for
assessing the resulting fully nonlinear turbulence.  The empirical
results of \cite{Hawley:2011} suggest that in the nonlinear regime
$Q_\phi \gtrsim 20$ and $Q_z \gtrsim 10$ are minimum values to describe
MRI-driven turbulence.

Shearing box results show that fully developed MRI turbulence has
certain average magnetic properties that also can be used
as diagnostics.  \cite{Hawley:2011} examined two such diagnostics,
the ratio of the Maxwell stress to the magnetic pressure, $\alpha_{mag} =
M_{R\phi}/P_{mag}$, and the ratio of the radial to toroidal magnetic energy,
$B_R^2/B_\phi^2$.  When suitably averaged over the computational domain
(as indicated by angle brackets, e.g., $\brbp$) these quantities appear
to approach specific values for well-resolved simulations.  The earliest
shearing box simulations \citep{Hawley:1995b} found that $\alpha_{mag}$ is
nearly constant from simulation to simulation.  \cite{Blackman:2008}
arrived at a similar conclusion across
a large sample of published unstratified shearing box results.
\cite{Hawley:2011} found that $\brbp$ increases
with $\Qz$ and seems to approach a value of $\sim 0.2$ at high $\Qz$.
Although $\brbp$ is related to $\alpha_{mag}$, they are not equivalent;
$\alpha_{mag}$ measures the correlation between 
$B_R$ and $B_\phi$, whereas $\brbp$ is a ratio of average energies,
with no dependence on correlations between the two components.

Halfway between the global disk and the shearing box is the {\it
cylindrical disk}, i.e., global models that do not include the vertical
component of gravity.  These are viewed as a model of the equatorial
region of a disk on scales less than $H$.  A study undertaken by
\cite{Sorathia:2012} computed a set of cylindrical disk models
using the {\it Athena} code, which has recently been extended to
cylindrical coordinates \citep{Skinner:2010}.  \cite{Sorathia:2012}
studied three different initial field topologies: sinusoidally varying
vertical field, net vertical field and net toroidal field.  They achieved
very high resolution, consistent with well-resolved shearing boxes, by
restricting the radial and vertical domain in the problem.  They also
used a numerical technique called ``orbital advection'' that takes
into account the background orbital motion by locally boosting into a
co-rotating frame.  This removes the orbital motion from the Courant
limit, increasing the timestep size.  This step may also reduce the
numerical diffusion that would arise from advection through the grid
due to the orbital motion.  Their fiducial grid has a radial domain from
$[1,4]$, a vertical domain $[-0.4,0.4]$ and a full $2\pi$ $\phi$ domain,
spanned by $480\times 1920\times 128$ zones for $\Delta R = \Delta z =
2\Delta \phi = 6.25\times 10^{-3}$.  The sound speed is set so that the
vertical domain is 4 isothermal scaleheights at $R=2$, giving a resolution
of 32 zones per $H_2$.  From these simulations they conclude that the most
important indicator of well-resolved MRI-driven turbulence, regardless of
initial magnetic field topology or strength, is the magnetic tilt angle,
$\theta_B = \arcsin (\alpha_{mag})/2 \approx 13^\circ$ \citep{Guan:2009};
this corresponds to $\alpha_{mag} \approx 0.44$.  They further stress the
importance of an effective resolution of $H/\Delta z > 32$, the use of
comparable grid zone size in all three dimensions, and the importance
of azimuthal resolution, recommending $R\Delta \phi \le 2\Delta z$.
They found, as suggested by \cite{Hawley:2011}, that higher $Q_\phi$
values ($\ge 25$) can compensate for lower $Q_z$ values, while $Q_z \ge
10$--15 can compensate for $Q_\phi \approx 10$.

The purpose of the present work is to extend these local-to-global
investigations using a systematic study of resolutions in full global
simulations.  We will test whether the diagnostics developed to date
continue to measure a simulation's quality, and attempt to bring a global
model toward adequate resolution.  We will compare different initial
magnetic field topologies and strengths and, since numerical accuracy
is not solely a question of grid zones, also some simple variations in
the numerical algorithm.

In \S2 of this paper we present the numerical model we will examine
and discuss some diagnostic measures.  \S3 presents a resolution study
for one specific fiducial initial magnetic field configuration. In \S4
we examine variations on both this base model and on the algorithmic
choices made within the {\it Athena} code.  Discussion and conclusions
are presented in \S5.

\section{Numerics and Diagnostics}

The simulations in this paper are all carried out using the {\it Athena}
code \citep{Stone:2008} using version 4.1 which incorporates
cylindrical coordinates, $(R,z,\phi)$, as developed by 
\cite{Skinner:2010}.  {\it Athena} solves the equations
of MHD in conservative form (mass, momentum, total energy and magnetic
flux) using a flux-conservative scheme.  The principal components of
{\it Athena} include the overall integration scheme, the
reconstruction algorithm within a zone, and the flux solver.  Except
where noted, we use the directionally unsplit CTU integrator of
\cite{Colella:1990}, third-order piece-wise parabolic reconstructions
\citep{Colella:1984}, and the HLLD flux solver of \cite{Miyoshi:2005}.

Since the aim of this study is to explore convergence in global
simulations, and high resolution is more easily attained by limiting the
domain size and evolution time required, we choose to study a region that
is close to the black hole.  The initial condition is based on 
the GT4 torus from \cite{Hawley:2000}
which was subsequently studied using a pseudo-Newtonian potential,
$\propto 1/(r-r_g)$, by \cite{Hawley:2001}.  That previous work used units
in which $r_g = 1$, and here we use units with $r_g =2M$.  This places the
innermost stable circular orbit (ISCO) of the pseudo-Newtonian potential
at $6M$. The computational
domain is a cylindrical annulus running from $R=4M$ to $R=44M$ and from
$-15M$ to $15M$ in $z$.  The $\phi$ domain spans $\pi/2$ in radians;
the size of the $\phi$ domain can influence the outcome of a simulation
\citep{Beckwith:2011,Flock:2011}, but for the present study we hold
that fixed.  We place outflow conditions on the $R$ and $z$ boundaries and
periodic boundary conditions in the $\phi$ direction.  The outflow
boundary condition is the same as used in \cite{Sorathia:2012}. There
are four halo, or ``ghost'' zones at each boundary and these are given
the same conserved state values as the last physical zone on the grid.
If a momentum vector is directed into the grid, it is reset to zero and
the total energy adjusted appropriately.

The initial torus has a pressure maximum at $R=20M$ and an inner boundary
at $R_{in} = 11M$, with an angular momentum distribution parameter
$q=1.68$, where $\Omega \propto R^{-q}$.  The closer the angular momentum
distribution is to Keplerian
($q = 1.5$), the thinner the initial torus will be.  We use an adiabatic
equation of state, $P = K\rho^\Gamma$ with $\Gamma = 5/3$, and $K =
0.0034$.  
Rather than using a scale height based on the isothermal sound speed,
we use the density scale height,
\begin{equation}
 H = {\int \rho |z| dz \over \int \rho dz} ,
\end{equation}
as a measure of disk thickness.  Using this, the
initial thickness of the torus at the pressure maximum is $H/R = 0.07$.
This definition of $H$ results in a smaller value than would be
obtained by using the sound speed at the equator; $H_2$ would be
larger by 1.9, for example.  In the absence of additional physics such as
a cooling function or self-gravity, the density units are arbitrary
and set by the choice of $K$.  The initial total mass is 788 in
code units (assuming a domain that spans the full $2\pi$ in $\phi$).
For the pseudo-Newtonian potential the Keplerian frequency is $\Omega =
\left(R^{1/2}[R-2]\right)^{-1}$, giving an orbital period
at the ISCO of $61.5 M$ and $506M$ at the torus pressure maximum.
The initial torus is seeded with random pressure fluctuations at the
1\% level.

The torus contains a weak (subthermal) initial magnetic field.
In \cite{Hawley:2001}, the initial magnetic is a nested dipolar loop
configuration (referred to here as a ``one-loop" configuration) where
the vector potential is set to be $A_\phi = C (\rho - \rho_{cut})$,
with $\rho_{cut}=0.25 \rho_{max}$, the maximum density at
the pressure maximum.  An alternative field topology consists of two
loops generated by toroidal currents of different signs (the ``two-loop"
configuration).  For this we use the the vector potential function of
\cite{Shafee:2008},
\begin{equation}
A_\phi = C \left[ (\rho - \rho_{cut}) r^{0.75}/\rho_{max}\right]^2 
\sin\left[\ln (r/S)/T\right]
\end{equation}
where $r$ is the spherical coordinate radius, $\rho_{cut}$ is set at 20\%
of the density maximum (thus confining the initial field to well within
the edge of the initial torus), $S = 1.1~R_{in}$, and $T=0.16$.  

The vector potential establishes the field geometry and the relative
strength of the field components throughout the torus, but the overall
field strength can be set by whatever normalization is desired.
The main requirement is that the field should have resolved MRI
wavelengths that fit within the torus.  We normalize the the initial
magnetic field strength using an average plasma $\beta$ value, $2 \langle
P\rangle/\langle B^2\rangle$.  Several initial $\beta$ values will be
examined.  We will find, however, that a very useful way to characterize
the field strength is with reference to the initial quality factors,
e.g. $Q_z$.  This immediately indicates the field strength relative to
the grid resolution within the torus, facilitating comparisons
of quite different simulations.

Using the results of \cite{Hawley:2011}
we can estimate the number of grid cells required to achieve certain
values of the $Q$ parameters in the evolving torus.  
For the vertical direction the number of cells per $H$ is estimated to be
\begin{equation}\label{eqn:qz}
N_z \simeq 16 \left(\beta/100\right)^{1/2} 
\left(\langle v_A^2\rangle/\langle v_{Az}^2\rangle\right)^{1/2}
\left(Q_z/10\right)
\end{equation}
where the ratio of the Alfv\'en speeds reflects the fraction of the
total magnetic energy in the vertical component alone, typically $\sim
0.01$--$0.1$.  Assuming that $\beta \sim 10$ in the turbulent phase, and
that the vertical field energy will be $\sim 0.05$ of the total magnetic
energy, condition (\ref{eqn:qz}) requires 21 vertical grid zones per $H$.

Adequate resolution of the toroidal MRI in a full $2\pi$ simulation is
estimated to require
\begin{equation}\label{eqn:qy}
N_\phi \simeq 1000~(0.1~R/H)~(\beta/100)^{1/2}~(Q_\phi/10)
\end{equation}
azimuthal cells.  Again, assuming $\beta \sim 10$ and $H/R \sim 0.2$
(the simulated torus will thicken due to disk heating), the $Q_\phi$
condition requires 75 $\phi$ grid zones in a $\pi/2$ domain.  At the
pressure maximum ($R=20M$) this requires $\Delta z \simeq 0.2$ and
$\Delta\phi = 0.021$ (corresponding to $R\Delta\phi = 0.42$ at the
pressure maximum).  A representative grid resolution would then be
$\sim 210\times 75 \times 160$ zones in $(R,\phi,z)$, not too-large
a number.  On a cylindrical grid, however, assuming a constant $H/R$,
the resolution requirement on $\Delta z$ increases as $R$ decreases.
A $\Delta z = 0.06$ is required to satisfy the $Q_z$ criterion at the ISCO
location of $R=6M$,which would increase the required number of vertical
zones to 500.  Since the proposed condition on the radial grid is simply
that the cell aspect ratio be on order unity, a comparable number of $R$
zones will be used.

To analyze the simulations, data is output once every ISCO orbital
period ($=61.5M$).  To obtain a rough overall measure of a simulation's
evolution, global totals such as mass, total energy, and components
of the magnetic and kinetic energies are computed every tenth of an
orbit at the ISCO.  Integrations over radial shells give azimuthally
and vertically averaged values at each radius.  For example, the mass
in a radial shell is
\begin{equation}
m(R) = \int \rho R d\phi dz ,
\end{equation}
and the accretion rate is
\begin{equation}
\dot M(R) = \int \rho v_R R d\phi dz .
\end{equation}

Since the focus is on the properties of the disk itself, most averages
will be density-weighted.  For example, $Q_\phi$
and $Q_z$ are intended to measure the functioning of the MRI-driven
turbulence within the disk, not in the corona where the Alfv\'en
speeds are much higher.  Thus they are computed as a function of
radius using
\begin{equation}
\langle Q_z \rangle(R) = { \int Q_z \rho R  d\phi dz \over
                        \int \rho  R d\phi dz }.
\end{equation}

On a more mundane note, the choice of grid resolution is also influenced
by the requirements of parallel domain decomposition, the number of
computational nodes available in a system, and the number of cores per
node.  The highest resolution simulations were run on the {\it Kraken}
system at the National Institute for Computational Science (NICS) which
has 12 cores per node.  For {\it Kraken}, maximum efficiency requires
use of domain decomposition in multiples of 12, whereas powers of two were
best for the University of Virginia Astronomy Department's {\it Hyades}
cluster,which was used for the lower resolution simulations.

\section{Resolution Study}

In the present study we choose one simulation to be the ``fiducial
simulation'' and vary the conditions, both physical and numerical,
relative to that benchmark.  Here, the designated fiducial simulation uses
an evenly spaced grid of $256 \times 64 \times 252$ zones, with $\Delta
R = 0.156$, $\Delta z = 0.117$ and $\Delta \phi = .025$.  By the $Q$
criteria (\ref{eqn:qz}), this resolution should produce a turbulent disk
that has $Q$ values close to, but just below, the thresholds of $\Qz =
10$ and $\Qp = 20$.  For a resolution test we double and halve the size
of the $R$ and $z$ zones compared to the fiducial grid.  The models are
labeled Twoloop-128 for the $128\times 64\times 128$ grid, Twoloop-256
for the fiducial $256\times 64\times 256$ grid, and Twoloop-512 for
the $512\times 120\times 512$ grid.  These are run to $\sim 10^4 M$
in time, or about 160 ISCO orbits.  Although this might seem to be a
limited evolution time (and in some respects it is), it is nevertheless
sufficient for most of the mass to be lost off the grid.  (For comparison,
the simulation of \cite{Hawley:2001} ran to only $3000 M$ in time.) The
fiducial simulation uses the two-loop initial field configuration with
an average initial magnetic field strength $\beta =1000$.  For the
two-loop configuration there are three regions of significant initial
vertical field:  near the disk inner edge, at the pressure maximum,
and at the disk outer edge.  These regions have maximum initial $\Qz$
values at the equator, reaching 3, 6, and 5.5 for the Twoloop-256 model;
the other two resolutions have twice or half those values.

We list these three simulations and all the subsequent variations
(considered in \S 4) in Table~\ref{table:list}.  In addition to the
two-loop model run at three resolutions, we run a set of two-loop
models with the same grid as the fiducial run, but different initial
field strength.  We also consider the fiducial run with algorithmic
variations: an alternate flux solver and second-order zone reconstruction.
Next we carry out simulations with alternative initial field topologies:
the dipolar loop configuration (one-loop) and a toroidal initial field.
Several resolutions are used for these alternative field topologies.
The table lists the number of grid zones and the grid zone sizes for
these simulations; since {\it Athena} currently does not allow for
variable grid zone size, this is fixed across the domain.  Any key
property of the simulation is noted in the column labeled ``Comment."
The number of vertical zones per scale height, $H/\Delta z$ is measured
at the initial pressure maximum, where $H/R = 0.07$, and is given for
reference.  By the standards of shearing box simulations, the zones per
scaleheight numbers are all low, but because $H$ increases as the disk
heats, this number will increase with time.

\begin{deluxetable}{lcccccl}
\tabletypesize{\scriptsize}
\tablewidth{0pc}
\tablecaption{Simulation List\label{table:list}}
\tablehead{
  \colhead{Name}&
  \colhead{$(R,\phi,z)$}&
  \colhead{$\Delta R$}&
  \colhead{$\Delta \phi$}&
  \colhead{$\Delta z$}&
  \colhead{$H/\Delta z$}&
  \colhead{Comment}
}
\startdata
Twoloop-128 & $128\times 64 \times 128$&0.313&0.0245&0.234& 6& $\beta=1000$\\
Twoloop-256 & $256\times 64 \times 256$&0.156&0.0245&0.117&13& Fiducial \\
Twoloop-512 & $512\times 120\times 512$&0.078&0.0131&0.059&25& \\
& \\
Twoloop-256w& $256\times 64 \times 256$&0.156&0.0245&0.117&13& $\beta = 4000$\\
Twoloop-256l& $256\times 64 \times 256$&0.156&0.0245&0.117&13& $\beta = 500$\\
Twoloop-256b& $256\times 64 \times 256$&0.156&0.0245&0.117&13& $\beta = 250$\\
& \\
Twoloop-256e& $256\times 64 \times 256$&0.156&0.0245&0.117&13& HLLE  \\
Twoloop-256s& $256\times 64 \times 256$&0.156&0.0245&0.117&13& 2nd Order  \\
& \\
Oneloop-128 & $128\times 64 \times 128$&0.313&0.0245&0.234& 6& $\beta=1000$\\
Oneloop-256 & $256\times 64 \times 256$&0.156&0.0245&0.117&13\\
Oneloop-512 & $512\times 120\times 512$&0.078&0.0131&0.059&25\\
Oneloop-320b& $320\times 120\times 240$&0.125&0.0131&0.125&10& $\beta=100$\\
& \\
Toroidal-64& $128\times 64\times 128$&0.313&0.0246&0.234& 6& $\beta_\phi=10$\\
Toroidal-128& $128\times 128\times 128$&0.313&0.0123&0.234& 6& \\
Toroidal-192& $256\times 128\times 192$&0.156&0.0123&0.156& 9\\
Toroidal-128s& $128\times 128\times 128$&0.313&0.0123&0.234& 6& 2nd Order\\
Toroidal-128w& $128\times 128\times 128$&0.313&0.0123&0.234& 6& $\beta_\phi=100$\\
& \\
Twoloop-512t& $512\times 128\times 512$&0.078&0.0123&0.059&25& $\beta=1000$\\
Twoloop-256t& $256\times 64 \times 256$&0.156&0.0245&0.117&--& re-grid\\
Twoloop-128t& $128\times 32 \times 128$&0.313&0.0490&0.234&--& $t=4600 M$\\
\enddata
\end{deluxetable}

\begin{figure}
\leavevmode
\begin{center}
\includegraphics[width=0.5\textwidth]{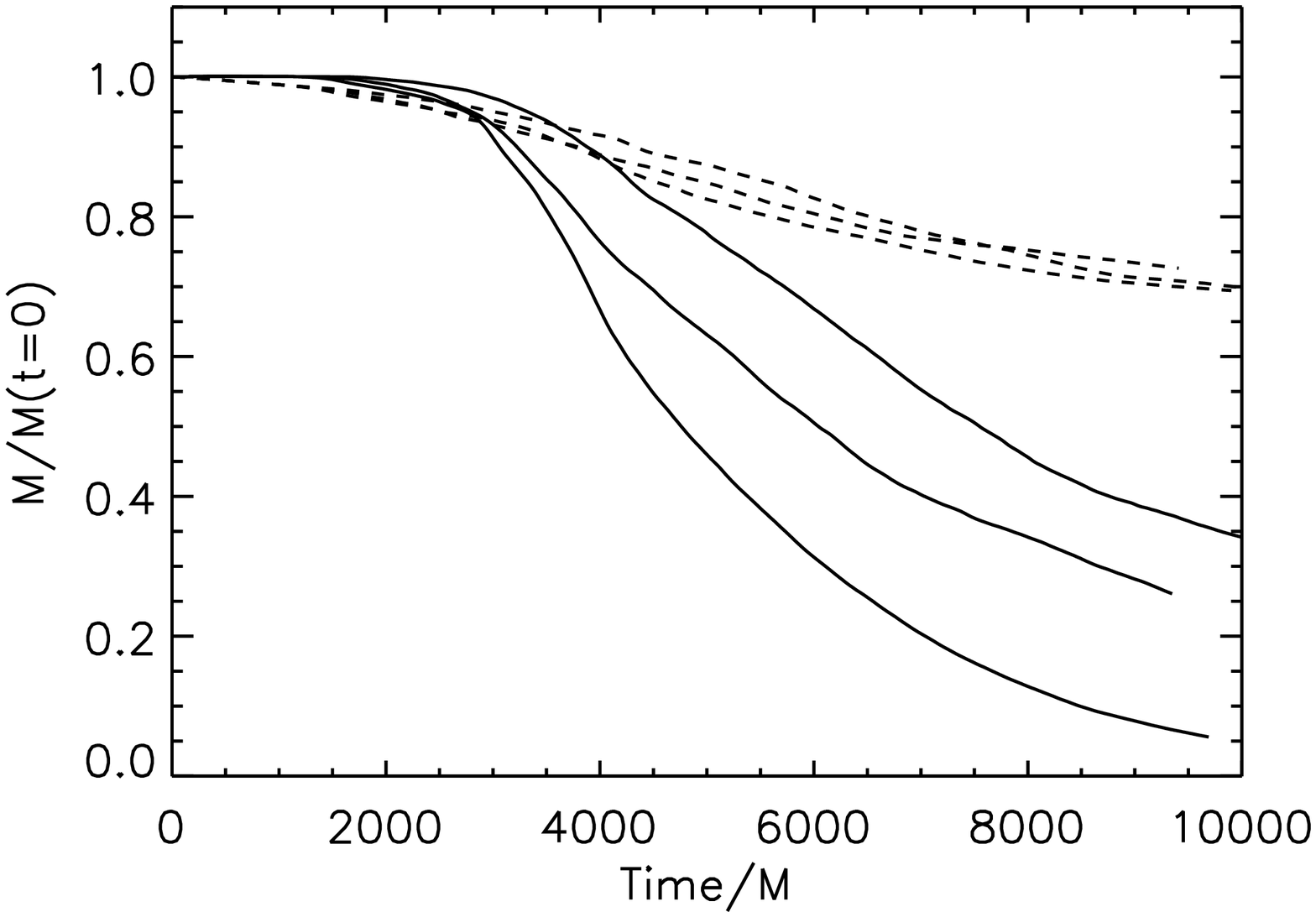}
\includegraphics[width=0.5\textwidth]{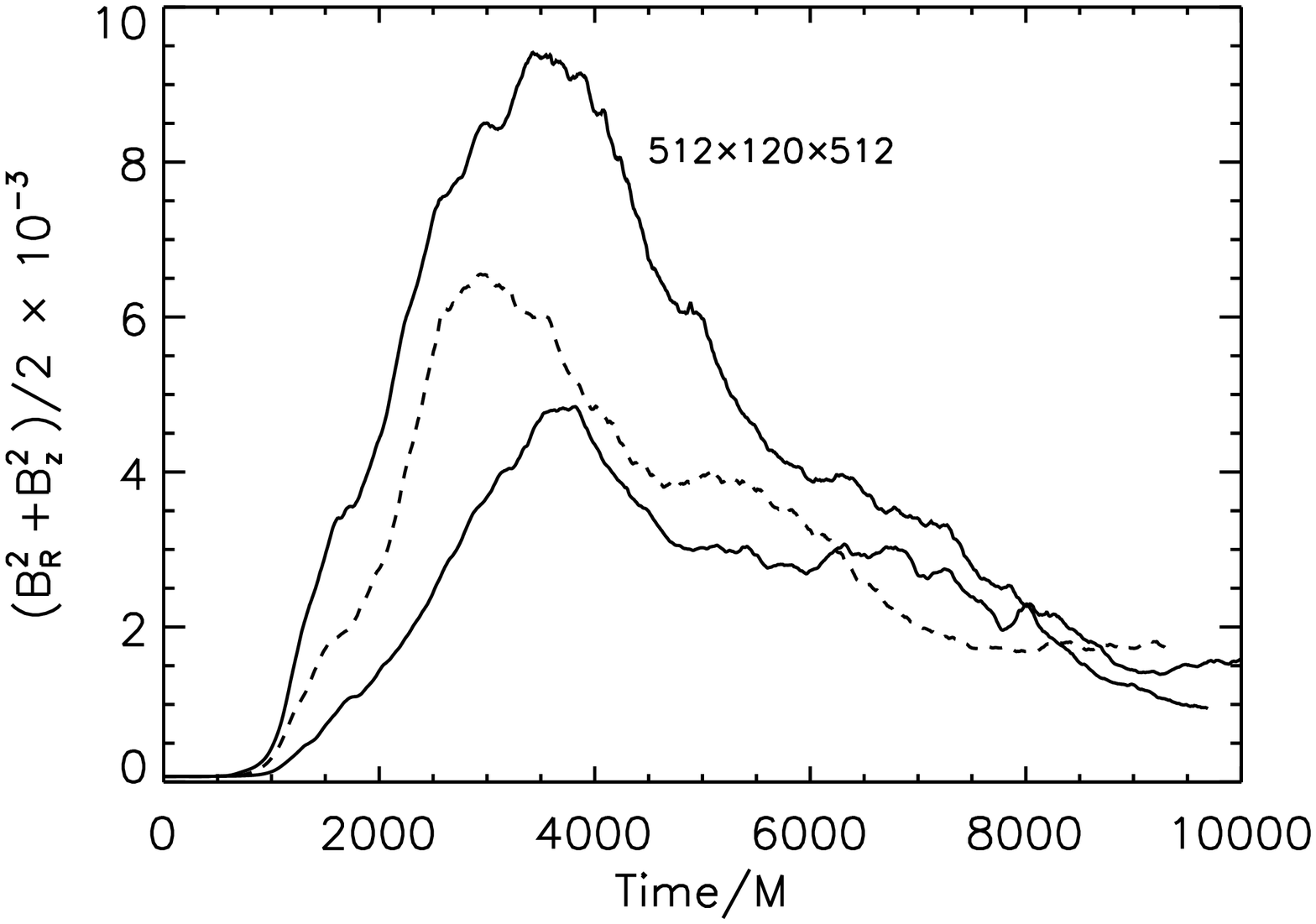}
\end{center}
\caption[]{(a) The total mass on the grid (solid lines) and the
accumulated mass accreted through the inner radial boundary (dashed lines)
as a function of time.  The $512\times 120\times 512$ model (labeled)
loses the most mass followed by the $256\times 64\times 256$ model and
finally the $128\times 64\times 128$ model.  The same ordering holds
for mass accreted, although the differences are not as large.  (b)
Time evolution of the poloidal field energy on the grid for two-loop
simulations of different resolution.  The $512\times 120\times 512$
model is labeled, the dashed line is the $256\times 64\times 256$
model, and the remaining line is the $128\times 64\times 128$ model.
} 
\label{fig:massevolve} 
\end{figure}

An overall sense of the evolution in the two-loop models is given
by the total mass on the grid, whose history is shown in
Figure~\ref{fig:massevolve}.  Each simulation shows a different
pattern in the development of the accretion rate, making
detailed comparisons between the simulations difficult.
Moreover, in all cases the initial condition is a torus of finite
mass.  There can be no ultimate long-term steady state in which
averaged properties can be defined because the mass on the grid
steadily diminishes; indeed, in all these simulations, the mass
at the end of the simulation is significantly less than at the
beginning.  This fact creates another problem of comparison if
one wishes to set these results beside those of
shearing box simulations, whose typical durations are
hundreds of orbits.  For example, \cite{Simon:2012}
regard the first 50 orbits of their shearing box simulations as the
``transient'' phase.  Here 50 orbits at the ISCO is $3000M$, and by
that point approximately 10\% of the initial torus mass is gone.

The evolution of all our simulations is qualitatively similar to what was seen in
\cite{Hawley:2001}.  A brief initial linear MRI growth phase leads
to turbulence throughout the disk, followed by an extended period
of evolution characterized by the steady decline in mass owing to
accretion and outward mass flow.
In Figure~\ref{fig:massevolve}a the dashed lines indicate the fraction of
the mass lost through the inner radial boundary.  The majority of the
mass leaves through the outer boundary.  By time $10^4 M$ in Twoloop-512,
31\% of the initial mass has accreted, 61\% of the initial mass has left
through the outer radial boundary, 6\% remains on the grid and a small
additional amount has left through the $z$ boundaries.  Net inflow
occurs primarily inside of a point just outside the initial disk
pressure maximum.  The location of $P_{max}$ moves to $R\approx 16M$
after $t=3000M$.  The accretion time at the initial pressure maximum
can be estimated as $\sim R/\langle v_R\rangle \sim 2\times 10^4 M$,
but the actual accretion time is much shorter.  Initially only 18\% of
the mass is located inside the pressure maximum, and by the end of the
simulation 31\% of the total mass has accreted, hence some of the mass
from outside of the initial torus pressure maximum has, in fact, accreted.
The plot shows that similar mass fractions accrete in the Twoloop-128
and Twoloop-256 run; indeed, for this quantity the major difference
between the different resolutions is the amount of mass lost through
the outer boundary.  As observed in \cite{Hawley:2011} and emphasized
in \cite{Sorathia:2012}, the rate of accretion is more rapid than would
be inferred from a simple estimate based on the average $\alpha$ value
and the radius of the initial torus pressure maximum; matter from
the initial torus quickly fills the inner disk as orbital shear creates
strong magnetic stress in the relatively concentrated field near its
inner edge.

Figure~\ref{fig:massevolve}b shows the evolution of the total poloidal
field energy as a function of time.  There are clear differences due
to resolution in the linear growth of the poloidal magnetic field.
The higher the resolution, the earlier and the faster the poloidal field
energy increases, and the larger it becomes.  Because MRI growth rates
are proportional to $\Omega$, field growth occurs at different rates
throughout the torus.  The inner part of the disk evolves first, with
turbulence developing by $t=1000M$, followed by the center of the disk
by $t=1500M$, and throughout the whole disk by $t=2500M$.  This greater
poloidal field energy corresponds directly to the larger stress and mass
accretion rates found in Twoloop-512 through $t=4000M$.  Beyond that point
the total field energy declines for all resolutions as mass and field
leave the grid.  This decline does not mean a reduction in magnetization
or a decay of the turbulence, however.  It is due to field lost off the grid
along with the mass.  The poloidal field energy per unit mass (Figure 1b
divided by Figure 1a) remains relatively flat with time for the two lower
resolution runs, and increases with time for the highest resolution model.

The initial growth of the toroidal field energy, $B_\phi^2$, is much less
dependent on resolution; it grows at the same rate for all
three simulations.  Toroidal field amplification occurs mainly through
azimuthal shear acting on the (initial) radial field, and this effect
is easily captured at even modest resolution. The peak toroidal field
energy value increases with resolution, however.   Beyond $t=4000 M$
the toroidal field energy per unit mass remains nearly constant for all
three models, but for Twoloop-512 this value is twice what it is in the
lower resolution simulations.

\begin{figure}
\leavevmode
\begin{center}
\includegraphics[width=0.5\textwidth]{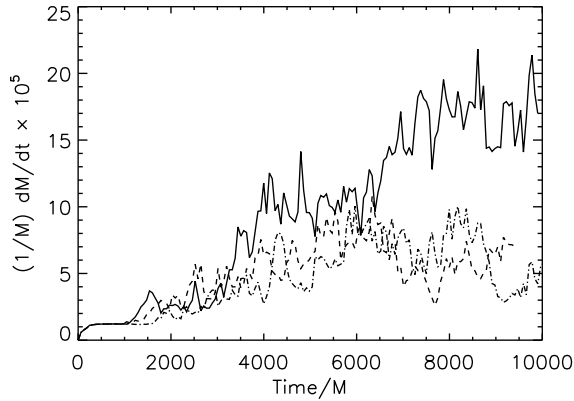}
\end{center}
\caption[]{Accretion rate divided by mass remaining on grid, $\dot
M/M$ as a function of time. The solid line is the
$512\times 120\times 512$ model, the dashed line the
$256\times 64\times 256$ model and
the dot-dashed line is $128\times 64\times 128$ model.  
} 
\label{fig:logmass} 
\end{figure}

We see that the overall evolution of the torus with time complicates the
interpretation of a number like mass accretion rate or field strength.
In Figure~\ref{fig:logmass} we compute the accretion rate at time $t$
as a fraction of the mass remaining on the grid at that time.  When
normalized in this manner, the highest resolution simulation stands
out as maintaining a robust accretion rate relative to the available
mass.

\begin{figure}
\leavevmode
\begin{center}
\includegraphics[width=0.4\textwidth]{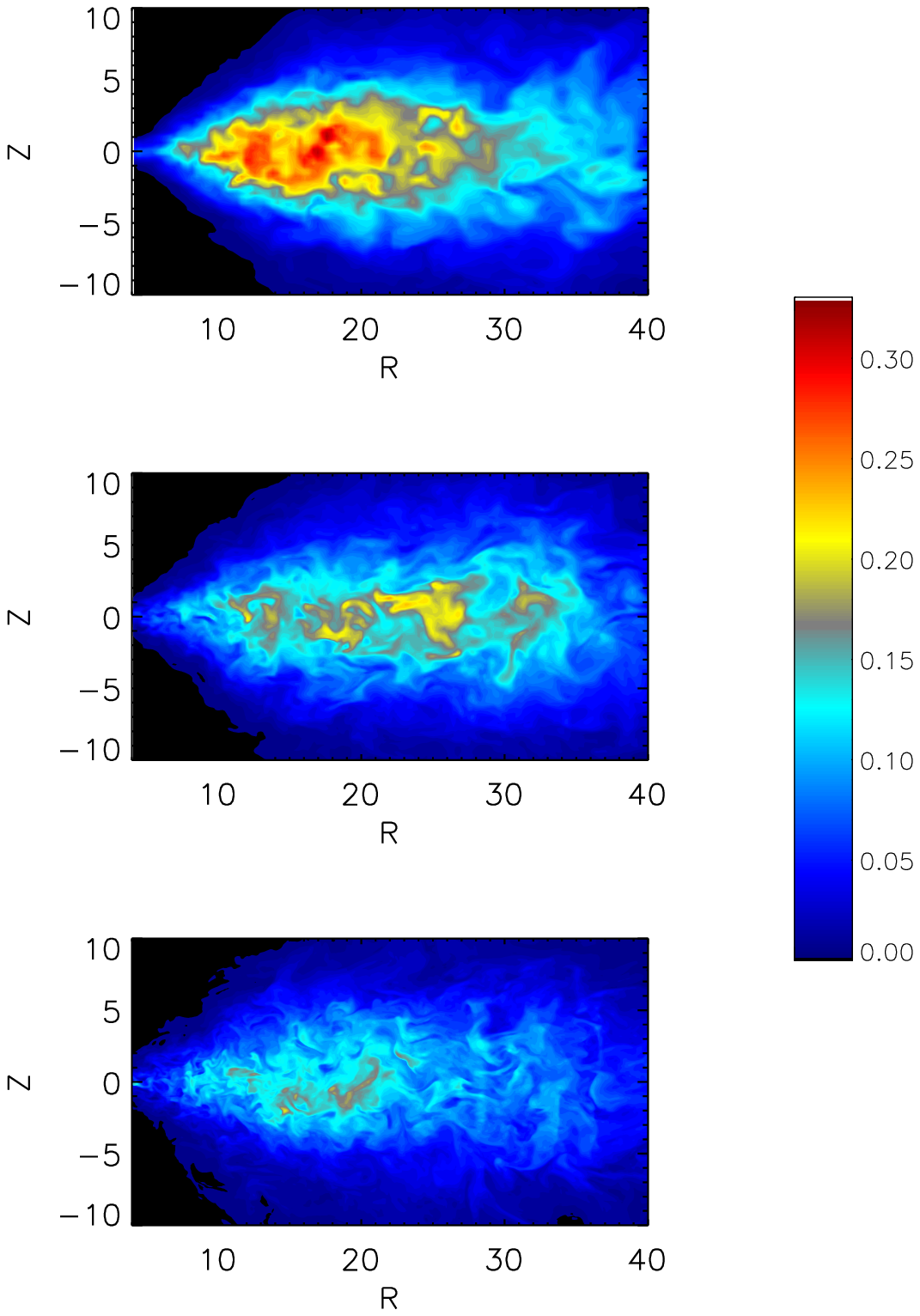}
\includegraphics[width=0.4\textwidth]{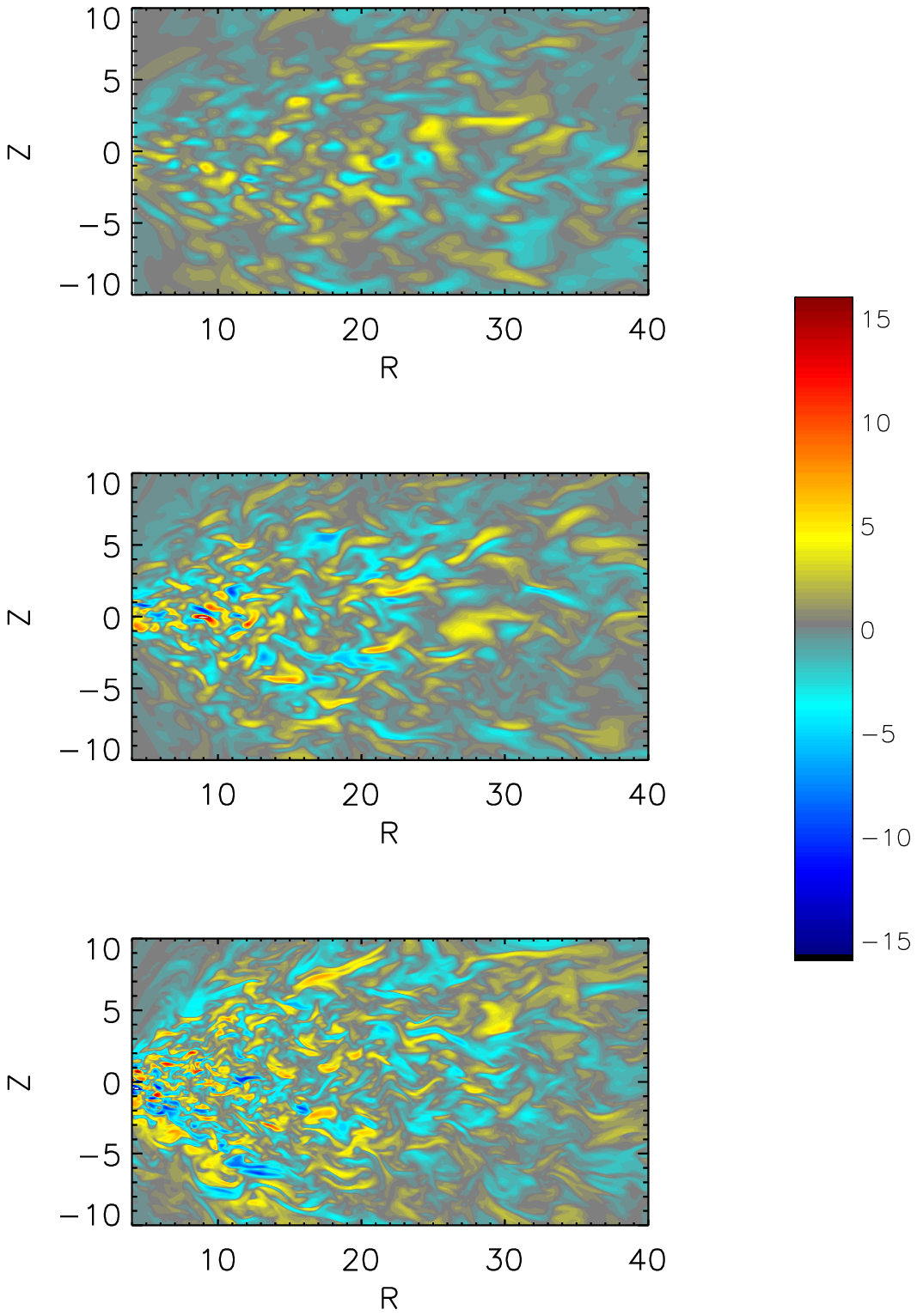}
\end{center}
\caption[]{(a) Density and (b) radial magnetic field 
for the initial two loop simulations with three
resolutions:  From top to bottom, Twoloop-128, Twoloop-256,
Twoloop-512.  The images show a single slice through the flow along
the $\phi=0$ plane, at $t=4920M$.  The density is
in units of the initial maximum density in the torus, and the magnetic 
field is in units of the initial maximum radial field strength.
}
\label{fig:twoloopfig}
\end{figure}

Figure~\ref{fig:twoloopfig} shows the three resolutions at $t=4920M$, when
the inner disk is accreting at a roughly constant rate.  The density plot
(left column) shows how increased turbulence and accretion have greatly
reduced the total mass in the highest resolution simulation compared
to the lowest.  The radial field plot (right column) suggests that
the larger scale fields are comparable from one resolution
to the next, but increased resolution brings stronger fields on small
scales and much more structure.

Table~\ref{table:results} gives diagnostic quantities for these runs,
time-averaged between $t=4000$--$6150M$, corresponding to the time after
the initial energy peak (Fig.~\ref{fig:massevolve}).  The accretion
rates are given both relative to the initial torus mass, $\dot
M/M(t=0)$, and to the current total mass, $\dot M/M$.  As suggested by
Figure~\ref{fig:massevolve}, the time-averaged $\dot M$ appears somewhat
insensitive to resolution, at least for this time period.  An examination
of accretion as a function of time finds that as resolution is increased,
accretion begins earlier and rises to higher initial values, but after
roughly $t \sim 5000M$ the lower resolution models have higher $\dot M
(t)$.  The reason is that the mass density is higher at late time in
the low resolution simulations; more of the torus mass remains to be
accreted.  The value of $\dot M/M$, however, shows a systematic increase
with resolution.

\begin{deluxetable}{lccccccccc}
\tabletypesize{\scriptsize}
\tablewidth{0pc}
\tablecaption{Simulation Results\label{table:results}}
\tablehead{
  \colhead{Name}&
  \colhead{$\dot M\times 10^5$}&
  \colhead{$\dot M/M\times 10^5$}&
  \colhead{$\alpha_{mag}$}&
  \colhead{$\Qz$}&
  \colhead{$\Qp$}&
  \colhead{$\brbp$}&
  \colhead{$\langle \beta\rangle$}&
  \colhead{$H/R$}&
  \colhead{stress $\times 10^{-6}$}
}
\startdata
Twoloop-512 &$4.8$&$10.2$&0.54--0.45&13--36&39--31&0.19--0.20
  &7.5--13&0.22--0.18&2.6--0.52\\
Twoloop-256 &$4.2$&$6.8$&0.50--0.46&4.5--15&19--16&0.13--0.18
  &8.1--13&0.20--0.18&2.0--0.57\\
Twoloop-128 &$4.5$&$5.9$&0.34--0.40&1.1--5.1&16-13&0.05--0.11
  &11--18&0.16--0.16&1.3--0.50 \\
& \\
Twoloop-256w&$2.7$&$3.1$&0.46--0.41&3.7--8.4&17--10&0.10--0.12
  &10.1--19&0.17--0.13&1.1--0.47\\
Twoloop-256 &$4.2$&$6.8$&0.50--0.46&4.5--15&19--16&0.13--0.18
  &8.1--13&0.20--0.18&2.0--0.57\\
Twoloop-256l&$4.7$&$10.0$&0.50--0.42&5.3--16&22--17&0.13--0.16
  &7.6--13&0.22--0.18&2.2--0.43\\
Twoloop-256b&$3.6$&$13.4$&0.56--0.44&9.2--26&30--22&0.19--0.20
  &4.7--10&0.31--0.23&2.5--0.38\\
& \\
Twoloop-256e&$4.3$&$5.9$&0.39--0.42&3.3--12&17--14&0.07--0.12
  &11--17&0.17--0.16&1.4--0.53\\
Twoloop-256s&$5.6$&$7.8$&0.43--0.39&3.4--13&19--17&0.08--0.11
  &8.1--14&0.18--0.17&2.5--0.52\\
& \\
Oneloop-128 &$3.2$&3.9&0.33--0.38&1.0--4.5&14-13&0.05--0.09
  &16--14&0.16--0.13&0.94--0.33 \\
Oneloop-256 &$2.5$&3.3&0.46--0.40&3.3--7.8&15--9.3&0.10--0.11
  &11--25&0.18--0.13&0.87--0.23\\
Oneloop-512 &$2.6$&3.7&0.52--0.43&8.1--22&28--21&0.17--0.17
  &9.7--18&0.19--0.15&1.20--0.43\\
& \\
Oneloop-320b&$2.8$&12.4&0.53--0.44&6--22&42--41&0.16--0.18
  &7.7--8.5&0.25--0.20&1.3--0.36\\
& \\
Toroidal-64 &$2.6$&4.1&0.27-0.34&1.0--2.8&14--10&0.04--0.07
  &17--29 &0.16--0.14 &0.62--0.25\\
Toroidal-128&$3.1$&1.1&0.38-0.42&1.4--7.1&33--32&0.07--0.13
  &13--15 &0.20--0.19 &0.82--0.34\\
Toroidal-192&$2.6$&4.8&0.47-0.44&2.4--9.2&30--24&0.11--0.15
  &12-26 &0.20--0.19 &0.86--0.24\\
Toroidal-128s&$1.7$&2.1&0.17-0.27&1.0--2.5&37--23&0.02--0.04
  &10--17 &0.18--0.13 &0.43--0.36\\
Toroidal-128w&$2.3$&5.0&0.33-0.39&1.2--4.3&28--23&0.05--0.10
  &17--25 &0.18--0.16 &0.52--0.23\\
& \\
Twoloop-512t&$4.0$&$ 8.0$&0.53--0.45&12--36&40--31&0.18--0.21
  &7.5--14&0.21--0.20&2.2--0.50\\
Twoloop-256t&$5.8$&$7.4$&0.51--0.45&4.4--14&19--14&0.14--0.18
  &8.2--20&0.21--0.20&1.6--0.34\\
Twoloop-128t&$6.3$&$7.6$&0.38--0.39&1.0--5.0&6.9-6.0&0.06--0.11
  &18--27&0.18--0.20&0.88-0.23 \\
\enddata
\end{deluxetable}

The other Table~\ref{table:results} values are density-weighted
averages on radial shells. For each diagnostic, two numbers are listed
corresponding to the value at the ISCO and at $R=20M$.  Because the grid
is cylindrical, the number of $z$ zones per $H$ will decrease inward
for any disk with a constant $H/R$.  $\Qz$ declines inward with
radius at a rate that is very nearly proportional to $R$.  This means
that the vertical Alfv\'en speed $v_{az}$ increases inward as $R^{-1/2}$
(eq.  \ref{eq:qzdef}).  $\Qz$ also increases slowly with time in the inner
part of the disk; the Alfv\'en speed increases with decreasing density
as the evolution proceeds.  A strong resolution effect is apparent: as
resolution decreases, $\Qz$ is reduced by more than can be accounted
for simply by the increase in $\Delta z$.

The value of $\Qp$ shows less dependence on $R$, but resolution does
influence its value.  The Twoloop-128 and Twoloop-256 models have the
same $\phi$ grid, so the increase in $\Qp$ from Twoloop-128 to Twoloop-256
is due to
the improved poloidal resolution.  The Twoloop-512 model increases the
number of $\phi$ zones to 120, and the increase in $\Qp$ is more than
can be attributed to the decrease in $\Delta\phi$.  Average magnetic
field strengths increase with resolution.

\begin{figure}
\leavevmode
\begin{center}
\includegraphics[width=0.7\textwidth]{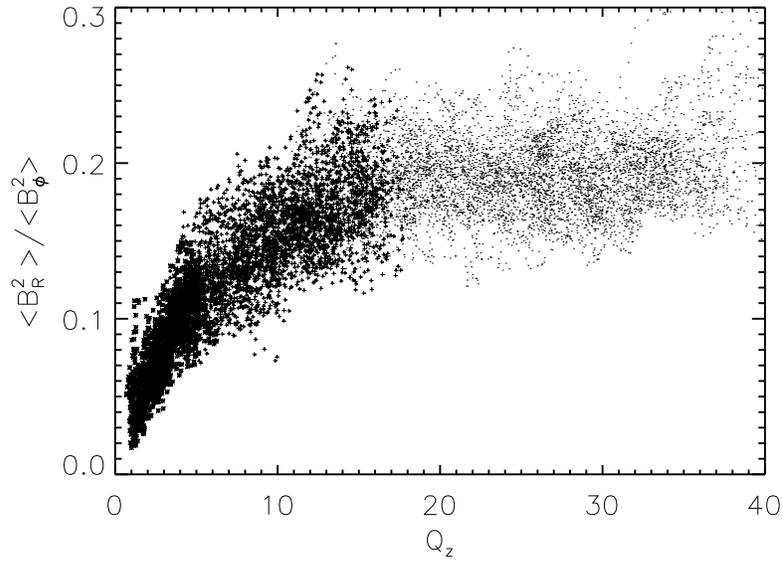}
\end{center}
\caption[]{
Values of $\brbp$ versus $\Qz$ for all radii between $6$--$20M$ and
for times from $t=4350$ to $6150M$.
Stars correspond to the Twoloop-128 run, plus signs to the Twoloop-256
run, and points to the Twoloop-512 run. }
\label{fig:brvsqz}
\end{figure}

\begin{figure}
\leavevmode
\begin{center}
\includegraphics[width=0.7\textwidth]{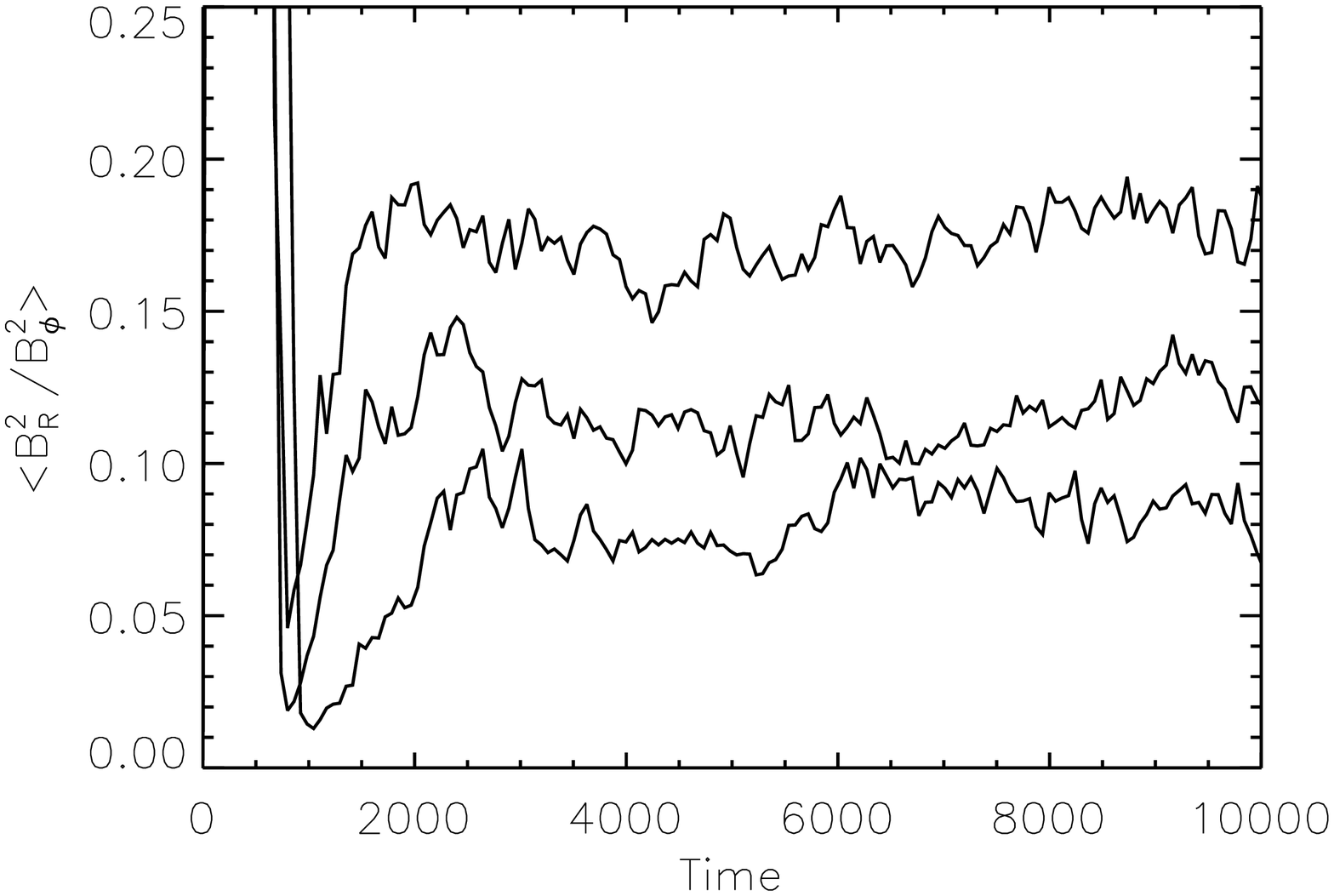}
\end{center}
\caption[]{Time evolution of $\brbp$ averaged between $R=6$ and $20M$
for the three resolutions of the two-loop simulations.
From top to bottom the curves correspond to Twoloop-512, Twoloop-256,
Twoloop-128.  Compare with Fig.~4 of \cite{Hawley:2011}.
}
\label{fig:brbyvst}
\end{figure}

The ratio of field energies, $\brbp$, also increases with resolution,
again, particularly near the ISCO.  Figure \ref{fig:brvsqz} examines the
relation between $\Qz$ and $\brbp$ by plotting the values computed for
individual radial shells for each radial grid zone
between $R=6$--$20M$, and for all individual data files between
$t=4000M$ and $6150M$.  The three resolutions sort themselves into
distinct regions of the graph with the value of $\brbp$ rising with increasing
$\Qz$ until $Q_z \approx 15$.  Within a given simulation, lower $\Qz$
values tend to correspond to smaller $R$; the spread of points represents the
inherent variability.  This figure improves upon Figure 14
of \cite{Hawley:2011}, where it was noted that $\brbp$ appeared to be
leveling off near a value of 0.2, but there were insufficient points with
high enough $\Qz$ to make that observation as conclusive as it is here.
Figure~\ref{fig:brbyvst} shows the time evolution of $\brbp$ averaged
between $R=6$ and $20M$ for the three runs.  Throughout the evolution,
the values of $\brbp$ remain distinct among the three resolutions.
A very similar behavior was seen in the shearing box simulations, shown
in Figure 4 of \cite{Hawley:2011}.

\begin{figure}
\leavevmode
\begin{center}
\includegraphics[width=0.7\textwidth]{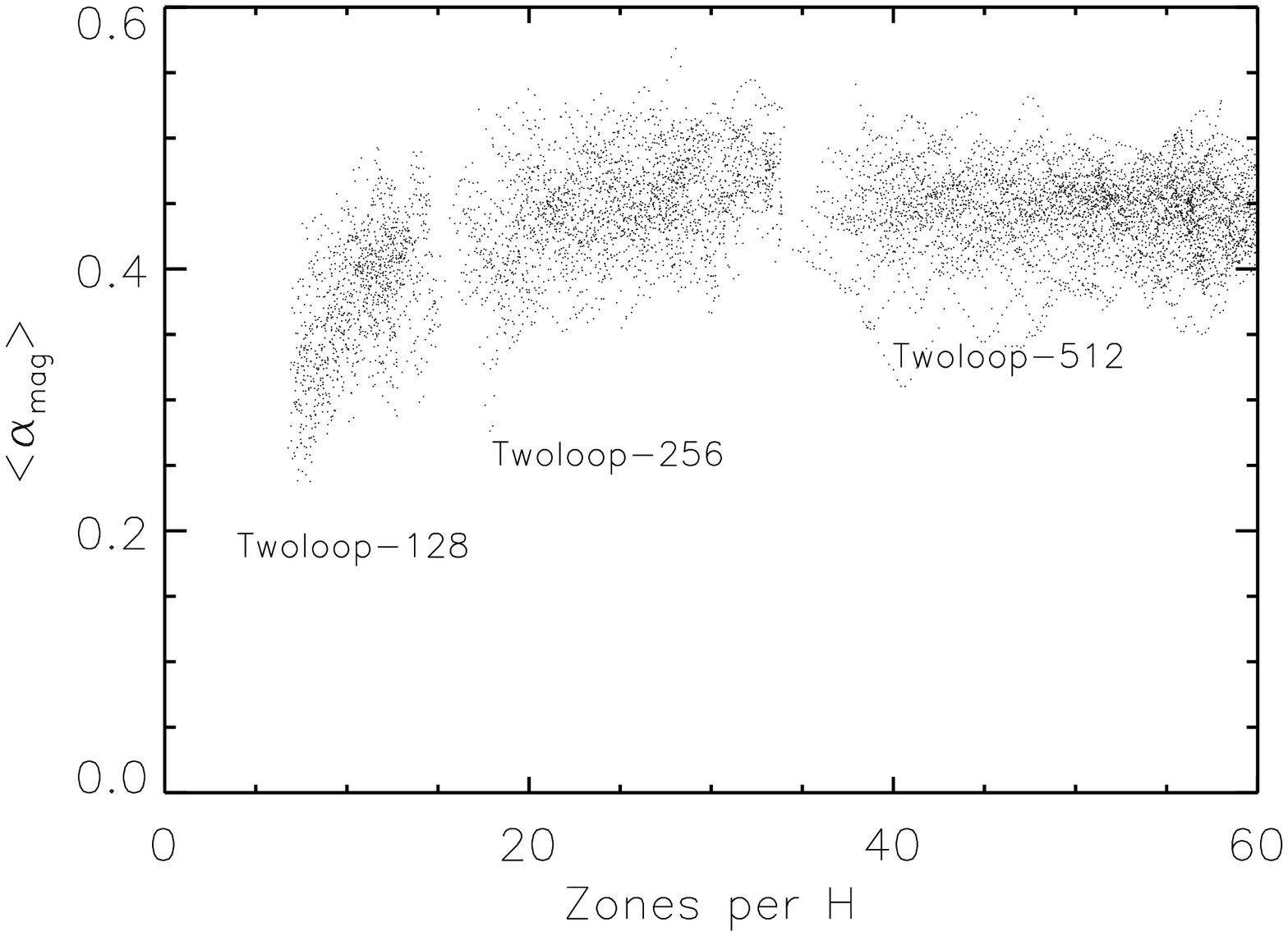}
\end{center}
\caption[]{
Values of $\alpha_{mag}$ versus number of grid zones per $H$ 
for all radii between $10$--$20M$ and for times from $t=4000$ to $6150M$.
The three different runs inhabit distinct regions of the plot, labeled
Twoloop-128, Twoloop-256 and Twoloop-512. The scatter shows the
relatively wide range of variation in $\alpha_{mag}$.  A systematic
decline is observed for resolutions less than 20 zones per $H$.
}
\label{fig:amagvsn}
\end{figure}

\begin{figure}
\leavevmode
\begin{center}
\includegraphics[width=0.7\textwidth]{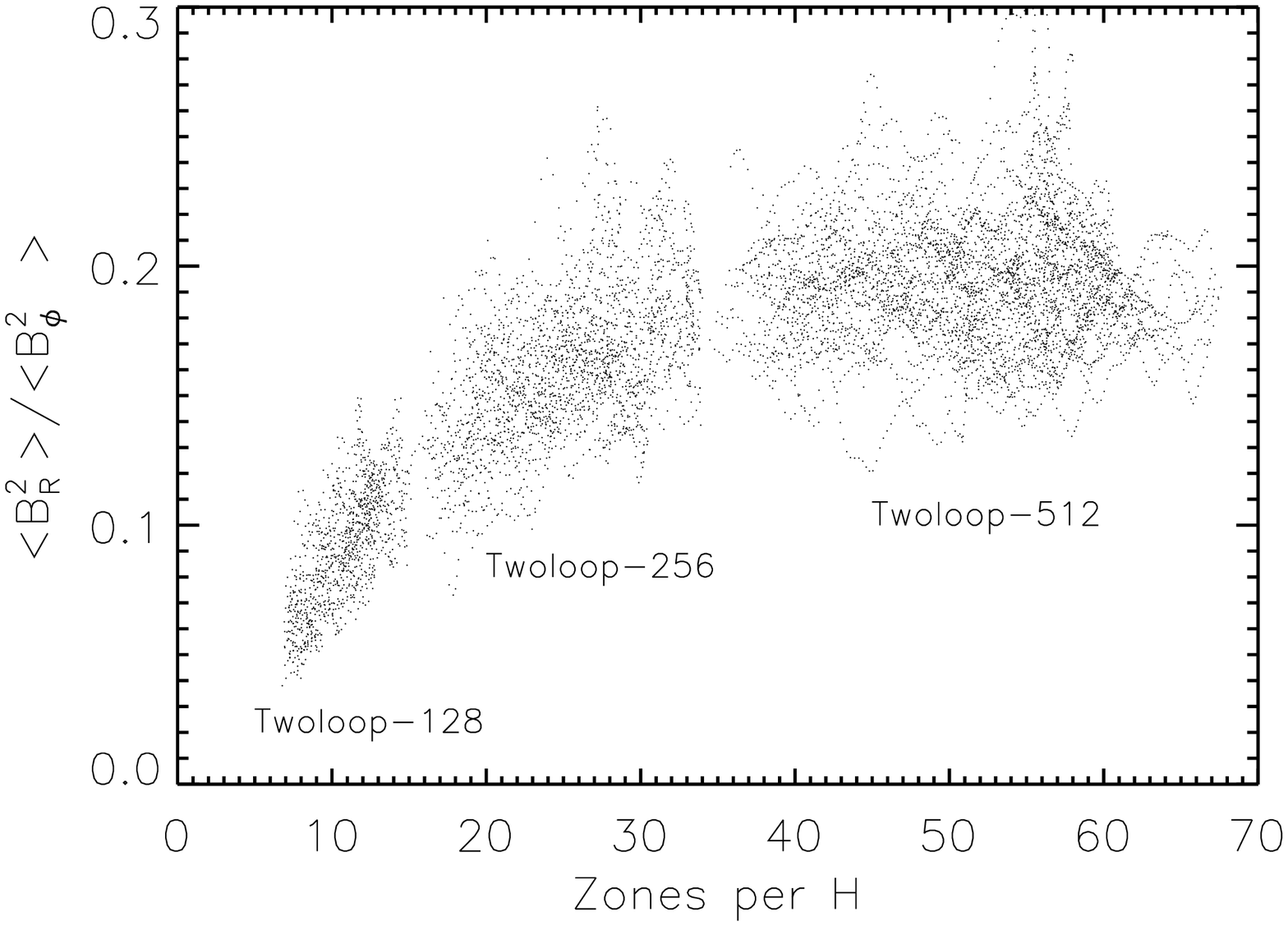}
\end{center}
\caption[]{
Values of $\brbp$ versus number of grid zones per $H$ 
for all radii between $10$--$20M$ and for times from $t=4000$ to $6150M$.
The three different runs inhabit distinct regions of the plot, labeled
Twoloop-128, Twoloop-256 and Twoloop-512. The scatter shows the
relatively wide range of variation in $\brbp$.  A systematic
increase with resolution is observed up to approximately 35 zones per $H$.
}
\label{fig:brbpvsn}
\end{figure}

Resolution also affects the value of $\alpha_{mag}$, although not
as dramatically.  For example, $\alpha_{mag}\sim 0.45$ for both the
Twoloop-256 and 512 runs at $R=20M$.  Figure~\ref{fig:amagvsn} plots
individual $\alpha_{mag}$ values for the three runs for all radii between
$10$--$20M$ and for times from $t=4000$ to $6150M$, and
Figure~\ref{fig:brbpvsn} plots $\brbp$ in the same way.  The data were
restricted to this radial range to avoid the region near the ISCO where
both quantities tend to increase owing to the transition from turbulence
to streaming inflow.  The three different resolutions separate nicely
into distinct regions due to two reasons:  increased resolution halves
the size of $\Delta z$, and the stronger magnetization in higher
resolution leads to greater heating and increased $H$.
From the first of these plots we see that $\alpha_{mag}$
is relatively constant (albeit with significant variation) with an
average value $\sim 0.45$ for resolutions greater than 20 zones per $H$.
Twoloop-128 exhibits a decrease in $\alpha_{mag}$ as the number of zones
drops below this value.  The plot shows that $\brbp$ increases
with resolution until leveling off at a value $\simeq 0.2$ when there are
roughly 35 zones per $H$.  This comparison suggests that while both of these
quantities are good indicators of resolution, the $\brbp$ criterion is
the more demanding.

At intermediate resolutions there is a general trend of increasing
$\alpha_{mag}$ with $\brbp$.  At high resolution, $B^2_r$ and $B^2_\phi$
become proportional across a wide range of values.  At lower resolution,
there is still a general proportionality, but $B^2_\phi$ tends to be
larger for a given value of radial field energy.  Lower resolution
results in less effective generation of poloidal field, although what
radial field there is remains correlated with $B_\phi$.

\begin{figure}
\leavevmode
\begin{center}
\includegraphics[width=0.7\textwidth]{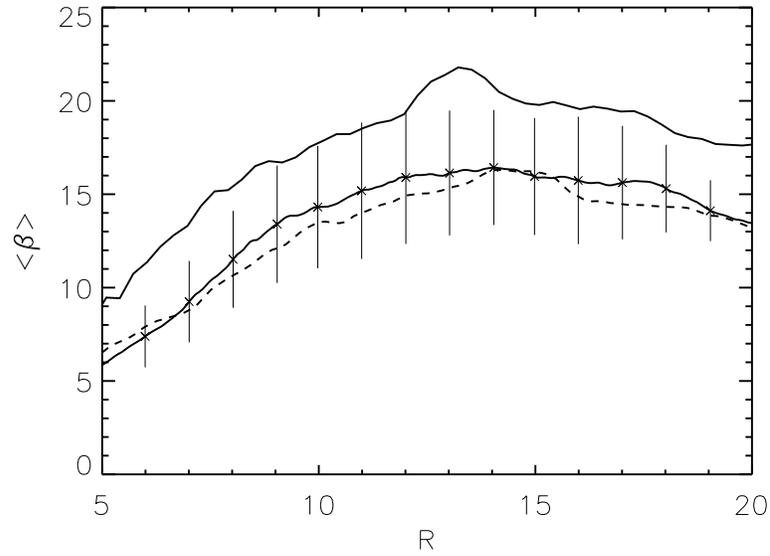}
\end{center}
\caption[]{The radial dependence of time-averaged $\langle
\beta\rangle$.  The time average is taken between $t=4000$--$6150M$.  
The top curve corresponds to the Twoloop-128 run,
the dashed curve to Twoloop-256, and the bottom solid curve to
Twoloop-512. The error bars on the Twoloop-512 curve show the standard
deviation of $\beta$ over the time interval.
}
\label{fig:betaplot}
\end{figure}

The density-weighted value of $\beta$, time-averaged between
$t=4000$--$6150M$ shows an increase in relative magnetization with
resolution.  Figure~\ref{fig:betaplot} plots the radial dependence of
time-averaged
$\langle \beta\rangle$ and reveals a systematic increase in magnetization
in going from the Twoloop-128 to the 256 zone model, but not much
difference between the 256 and 512 models.  Although the total magnetic
energy changes significantly over the simulation in response to mass
loss, the magnetization $\beta$ shows much less variation within the
inner disk.  The error bars in Fig.~\ref{fig:betaplot} show one standard
deviation in value over the averaging time for Twoloop-512.

\begin{figure}
\leavevmode 
\begin{center}
\includegraphics[width=0.7\textwidth]{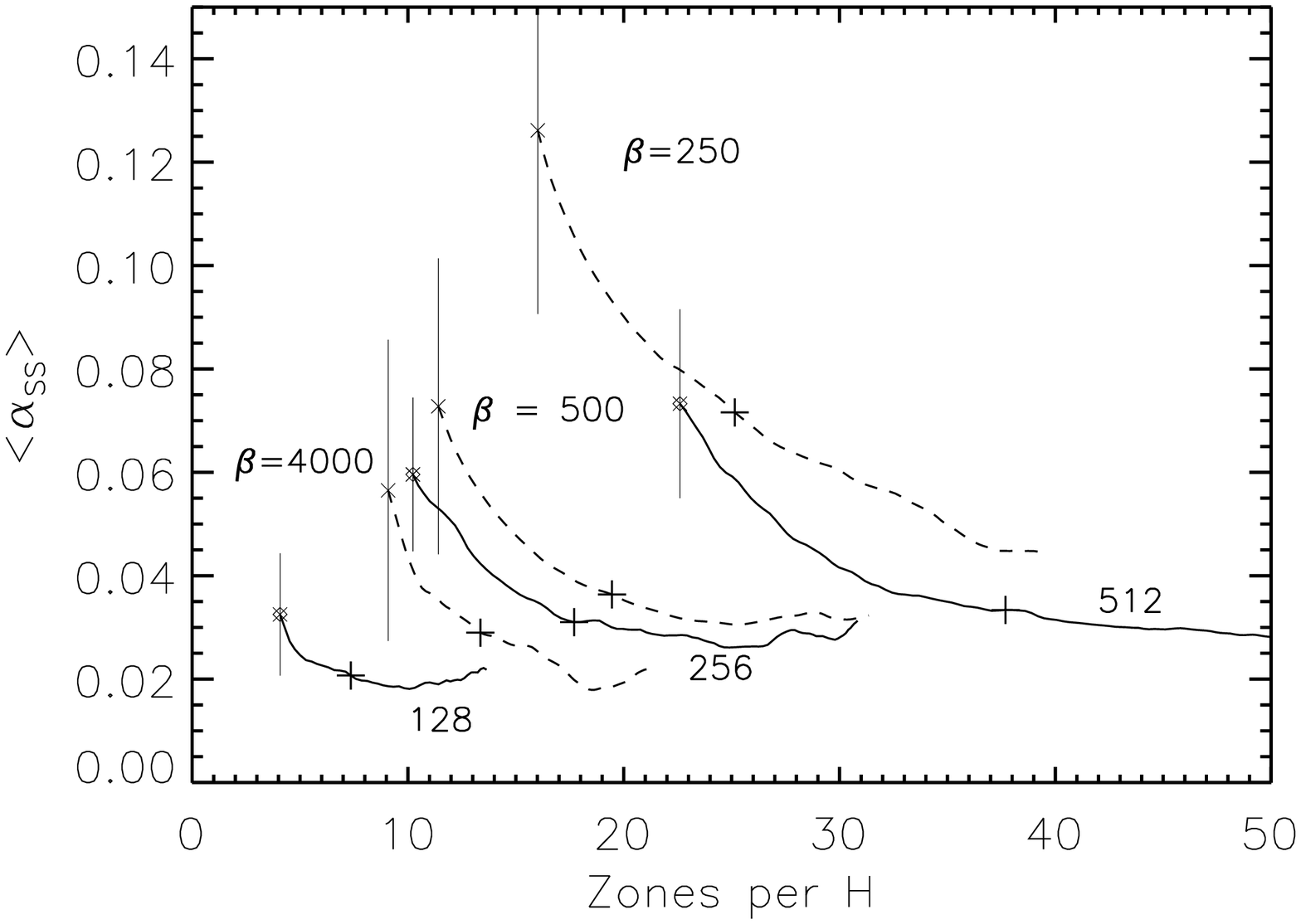}
\end{center} 
\caption[]{The density-weighted Maxwell stress parameter, 
$\alpha_{SS} = -B_R B_\phi / 4\pi P_{gas}$, time-averaged from
$t=4000$--$6150M$, versus number of zones per time-averaged scale-height
between $R=6$--$20M$ for several two-loop runs.  The solid lines are 
the three resolution experiments and the dashed lines are field strength
experiments labeled by the initial $\beta$ value.  Scale height increases
with radius and, since $\Delta z$ is fixed for each resolution, plotting
against number of zones per $H$ provides separation for the curves.
The error bar at the end of each curve corresponds to the standard
deviation of the stress over the time interval at the ISCO.  The
cross on each curve marks the value of $\alpha_{SS}$ at $R=10M$.
}
\label{fig:alphass}
\end{figure}

Figure~\ref{fig:alphass} is a plot of the time-averaged Shakura-Sunyaev
parameter $\alpha_{SS} = \alpha_{mag}/\beta$ as a function of zones per
scale height $H$ for the Two-loop runs, from the ISCO out to $R=20M$.
(Note that this definition of $\alpha_{SS}$ does not include the Reynolds
stress.)  The advantage in plotting $\alpha_{SS}$ instead of the code
value of the Maxwell stress is that the latter varies systematically
with mass loss due the overall evolution of the disk; $\alpha_{SS}$
shows much less secular variation with time.  Plotting against zones
per scaleheight provides separation for the curves;  disks that are
thicker or that have higher resolution will be shifted to the right.
Moving left to right along each curve corresponds to moving from smaller
to larger $R$.   The point marked with a cross on each curve corresponds
to the value at $R=10M$.  In order of increasing resolution these
values are $\alpha_{SS} = 0.02$, 0.033, and 0.034.  For Twoloop-512,
$\alpha_{SS} \approx 0.03$--0.04 between $R=10$--$20M$.  $\alpha_{SS}$
increases sharply near the ISCO as the gas pressure falls and
$M_{R\phi}$ increases.  This is an
example of a global effect that would not be seen in a shearing box model.
Although our focus is on the properties within the main body of the
disk, it is worth
noting that the stress increases rapidly inside the ISCO, and this
``plunging region'' stress increases systematically with resolution.
(Again, recall that with a cylindrical grid the number
of zones per scale height decreases inward when $H/R$ is constant.)

The next column of Table~\ref{table:results} gives the average disk
thickness, $H/R$, at the ISCO and $R=20M$.  Although the initial disk had
$H/R = 0.07$, it thickens considerably.  Disk turbulence is sustained
by the continuing action of the MRI, and the turbulent magnetic and
velocity fluctuations are dissipated at the grid scale.  {\it Athena}
uses an energy conserving algorithm; numerical losses in kinetic and
magnetic energy are converted into heat.  The resulting increase in $H$
improves the nominal measure of resolution quality, namely the number of
$z$ zones per $H$.  At $t=5474M$, for example, the three different
resolutions have $H/\Delta z$, of 64, 32, and 14 zones at $R=20M$.
At the ISCO, $H$ is smaller and the zones per scale height is reduced
to 22, 10, and 4.

Finally, the last column in the Table gives the density-weighted Maxwell
stress.  Not surprisingly, given the zones per scale height there,
the biggest relative difference between resolutions is near the ISCO.

How do the outcomes compare with the predicted zone requirements as
expressed in Eqns.~(\ref{eqn:qz}) and (\ref{eqn:qy})?  Using the time-averaged
values for $\beta$, $H/R$ and the components of the magnetic energy, the
number of $\phi$ zones needed to achieve at least $\Qp = 10$, $N_\phi$,
ranges from 31--50, 35--50, and 51--66 between the ISCO and $R=20M$ for
the Twoloop-512, 256 and 128 models respectively.  All three models are
adequately resolved in $\phi$ by this criterion, although Twoloop-128 is
only marginally so, and its value of $\Qp$ is the lowest.  The required
number of vertical zones per $H$, $N_z$, is another matter, however,
with computed values of 21--23, 28--24, and 57--34 for the Twoloop-512,
256 and 128 models.  At the ISCO, only Twoloop-512 model meets this
criterion, and Twoloop-128 doesn't even meet it at $R=20M$.
Thus, the number of zones required to maintain adequate resolution
throughout a simulation {\it decreases} with {\it increasing}
initial resolution.  The reason for this behavior is that simulations
with better-quality initial resolution achieve stronger field strengths
at saturation, whose demands on resolution are weaker, while simulations
with initially poor resolution create weaker fields, requiring finer
resolution to describe.  As a result, only simulations that are
well-resolved at the initial condition can hope to reach physical
saturation levels.

In summary, there are differences between all three resolutions studied
here.  In the initial evolution the magnetic field grows more rapidly
and to a higher amplitude with increasing resolution.  The initial
field strength is weak enough that the initial $Q_z$ values are
somewhat marginal for Twoloop-256, and clearly under-resolved
for Twoloop-128.  This is reflected in the early evolution, and
the level of magnetization reached at the end of initial growth period
influences what occurs subsequently.  The highest resolution simulation
has achieved the targeted $Q$ values.  Whether the simulation is
``converged'' at least in the sense that no further significant 
quantitative differences
would be seen with higher resolution, remains uncertain, but seems likely.  The plot of $\brbp$
versus $\Qz$ suggests that Twoloop-512 is adequately resolved and that
$\brbp \approx 0.2$ is the value achieved under those circumstances.
By the same criteria, Twoloop-256 is nearly adequately resolved and
Twoloop-128 is clearly under-resolved.  This is reflected in the relative
stress levels seen in the simulations.  Looking at the number of grid
zones per $H$ in the simulations, the results are consistent with the
shearing box conclusion that ``adequate resolution'' is achieved for $>32$
zones per $H$.

Despite being demonstrably under-resolved, the Twoloop-128 model
nevertheless evolves in a {\it qualitative} manner similar to the higher
resolution models, at least initially.  Low resolution simulations can
be usefully employed in surveys of disk evolution scenarios, but caution
is in order in assessing quantitative values, especially if confidence
in the simulation is based solely on the appearance of seemingly robust
accretion.

These conclusions follow from a particular problem configuration.  We
next consider the effects of variations on that configuration.

\section{Simulation Variations}

\subsection{Effect of Initial Field Strength}

Accretion simulations begin from an idealized initial condition usually
consisting of some well-ordered magnetic field configuration.  The long
term evolution of the resulting flow is driven by MRI-driven turbulence
that develops from this initial state.  Well-resolved ideal MHD stratified
shearing box simulations have found that MRI turbulence can be sustained
indefinitely and that the properties of the turbulence are independent
of the topology or strength of the initial field in the absence of a
net $B_z$ field through the box.  (The situation is more complex for
non-ideal MHD.  The study of
\cite{Rempel:2010} shows, for example, that finite
resistivity can cause zero net flux MRI-driven nonstratified shearing
box turbulence to gradually decay.) Compare the recent
high-resolution results from \cite{Simon:2011}, \cite{Davis:2010},
\cite{Guan:2011} and \cite{Simon:2012}, for example.  The key term here
is ``well resolved.''  Turbulence either fails to develop or dies out
if the initial field is too weak with respect to the resolution (i.e.,
low $Q$ value), or if there is insufficient range between $\lambda_{MRI}$
and the resistive scale in simulations where resistivity is included.

The {\it first} requirement for an adequately resolved simulation is
an adequately resolved initial field as indicated by $\Qz$ or $\Qp$.
To explore the effect of the initial field strength we repeat the two-loop
fiducial simulation with a weaker initial field ($\beta =4000$,
Twoloop-256w), and stronger fields ($\beta = 500$, Twoloop-256l;
$\beta =250$, Twoloop-256b).  The field strength is characterized in
terms of $\beta$, but for the purposes of resolution considerations it
is more appropriate to use $\Qz$.  With the two-loop field configuration,
the maximum initial $\Qz$ is found near the pressure maximum and is equal to 2,
4, 5.7, and 8 from the weakest to the strongest field.  As previously
remarked, the fiducial run is already below the standard criterion of $\Qz
> 6$, so the weakest field model will be at an even greater disadvantage.

Although the initial field topology is the same in all these runs,
the different field strengths lead to considerable differences in the
evolution.  The poloidal magnetic field grows due to the MRI and
all wavelengths longer than $\sim \lambda_{MRI}/\sqrt{3}$ are unstable.
While these wavelengths are unstable (e.g., those on order
$H$), the growth rate at a fixed wavelength longer than $\lambda_{MRI}$
is $\propto v_A$, and hence is small for
weak fields.  Both these factors retard the overall
field amplification and the onset of turbulence as $\beta$ increases. The
weakest field model, Twoloop-256w, reaches its peak poloidal field energy
at $t=4100M$.  As field strength increases, the peak occurs earlier:
the fiducial run peaks at $t=3000M$, Twoloop-256l at $t=2800M$, and
Twoloop-256b at $t=2700M$.  For the toroidal field, amplification is
primarily through shear acting on the radial field, $dB_\phi/dt \propto
-B_R \Omega t$, hence the growth rate of $B_\phi$ is proportionally
reduced for weaker fields.

Figure~\ref{fig:alphass} plots the time-averaged $\alpha_{SS}$ for
the Maxwell stress as a function of zones per scale height $H$ for the
two-loop runs.  The $\beta = 250$ model, Twoloop-256b, has a particularly
large Maxwell stress, apparently benefiting both from the larger
initial field strength and the thicker disk that develops due to greater
rates of heating.  From weakest to strongest initial field strength,
$\alpha_{SS}$ at $R=10$ (marked with a cross in Fig.~\ref{fig:alphass})
is 0.029, 0.033, 0.036 and 0.072.  Again, $\alpha_{SS}$ is
proportional to magnetization, $\beta^{-1}$.  The accretion rate as a
function of mass on the grid also increases within initial field
strength, consistent with the observed increase in $\alpha_{SS}$.

Table~\ref{table:results} shows that all the quality metrics improve
with increasing initial field strength.  Here the greatest difference
is between the weakest initial field, Twoloop-256w, and the others.
The values of $\brbp$ increase substantially, almost doubling from
Twoloop-256w to Twoloop-256b.  On the whole
Twoloop-256w resembles Twoloop-128.  The values of the initial $\Qz$
in both those simulations are equivalent.  This illustrates that the
question of adequate resolution is not solely one of grid zones, or even
of initial $\beta$.  Rather the ratios between the grid zone size and
the MRI wavelength---i.e., $\Qz$ and $\Qp$---are the best indicator,
not only for the initial field growth, but also for the subsequent evolution.

\subsection{Effect of Numerical Algorithm}

The outcome of a numerical simulation is not solely determined by the
grid resolution; the numerical algorithm itself can be very significant.
Algorithmic effects are particularly important for under-
or marginally-resolved simulations.  The optimal
choice of algorithm may not be clear-cut.  \cite{Kritsuk:2011}, for
example, carried out a detailed comparison of a number of MHD codes on
a particular turbulence simulation, and the results were qualitatively
similar, but with some variation in details.  No scheme emerged
as the clear winner.  \cite{Flock:2010} studied the effect of
numerical scheme and Rieman solver on the linear MRI in a
global disk and found that the proper treatment of Alfv\'en
characteristics in the Riemann solver is crucial.

Because {\it Athena} is designed with a choice of state reconstruction
schemes and flux solvers, it offers an opportunity to test what a
single algorithmic change has on the various physical and diagnostic
quantities on the same global disk problem.  By default we use the
third-order piecewise-parabolic reconstruction with the HLLD flux solver.
Using our fiducial run as a baseline we can compare simulations that use
second-order, piecewise-linear reconstructions and the HLLD flux solver
(Twoloop-256s), and a simulation with third-order reconstruction and the
HLLE flux solver (Twoloop-256e).  In principle, the use of second-order
reconstruction will lead to larger truncation error compared to the use
of third-order reconstruction.  As for the flux solvers, the HLLE scheme
uses only information from the fastest wave speeds; it is diffusive for
contact discontinuities.  The HLLD scheme corrects this by considering
the full family of wavespeeds, but at a cost of greater computational
complexity.  \cite{Beckwith:2011b} found that HLLD was approximately 1.5
times more accurate than the HLLE solver as measured in an Alfv\'en wave
convergence test.

\begin{figure}
\leavevmode
\begin{center}
\includegraphics[width=0.7\textwidth]{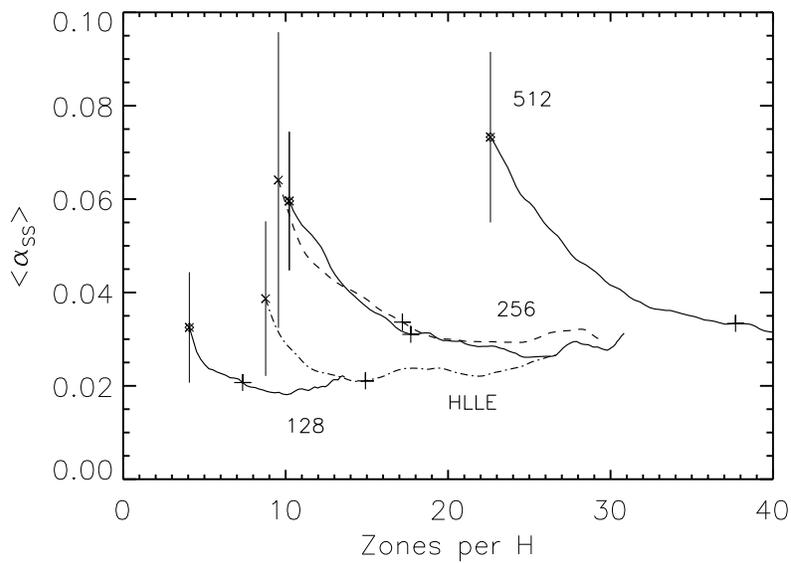}
\end{center}
\caption[]{The Shakura-Sunyaev for the density-weighted Maxwell
stress $\alpha_{SS} = -B_R B_\phi / 4\pi P_{gas}$ time-averaged from
$t=4000$--$6150M$, versus number of zones per time-averaged scale-height
between $R=6$--$20M$ for several two-loop runs.  The solid lines are
the three resolution experiments, the dashed line corresponds to the
model run with second-order reconstruction, and the
dot-dashed line is the model run with the HLLE flux solver.
The error bar at the end of each curve corresponds to the standard
deviation of the ISCO stress over the time interval.
}
\label{fig:hllestress}
\end{figure}

A comparison with the fiducial run shows that either algorithmic change
delays the initial linear growth phase and reduces the field energy
growth rate.  Use of the HLLE flux solver most reduces the growth rate.
Table~\ref{table:results} also shows some decrease in the quantitative
diagnostic values; a reduction in $\brbp$ is particularly noticeable in
both Twoloop-256s and Twoloop-256e.
Figure \ref{fig:hllestress} is the time-average value of $\alpha_{SS}$
as a function of zones per scale-height $H$ in the three resolution
simulations and the second-order and HLLE simulations.  The value of
$\alpha_{SS}$ at $R=10M$ is 0.021 for the HLLE simulation, the same as
Twoloop-128, and 0.034 for the second-order simulation, the same as
the fiducial run.  The HLLE simulation's effective resolution is 
comparable with the Twoloop-128 run, while the use of second-order
reconstruction has had only a minimal effect compared to the fiducial
simulation.  This result is not, however, universal.
In \S 4.4 we will see that the use of second-order interpolation has a
significant effect when the initial field is toroidal.

\subsection{One Loop}

The simulations with a two-loop initial field configuration found that
both resolution and initial field strength can affect the quantitative
outcome.  To test the impact of initial field topology, we next consider
three resolutions of an initial $\beta =1000$ one-loop field, labeled
Oneloop-512, Oneloop-256 and Oneloop-128.  For the one-loop field, the
vertical field is located at the front and back of the torus.  With this
$\beta$ the peak initial $\Qz$ is 2.5 for the Oneloop-256
simulation, well below the nominal threshold for adequate resolution.
The initial maximum $\Qz$ for Oneloop-512 is 5, comparable with the
initial $\Qz$ in Twoloop-256.

The evolution of a one-loop model is {\it qualitatively} similar
to a two-loop model, and the relative effects of resolution are
the same, but the one-loop models have lower overall magnetization.
The poloidal field energies and accretion rates are reduced compared to
the equivalently gridded two-loop models (see Table~\ref{table:results}).
The role of the toroidal field seems particularly important.
The toroidal field energy increases at the same rate for
all three resolutions---it grows due to shearing of the initial radial
field---but the lowest resolution model has the highest peak value.
This is due to a delay in the onset of turbulence caused by lower
resolution, which gives the  relatively unperturbed coherent radial
field a longer period of shear amplification.

\begin{figure}
\leavevmode
\begin{center}
\includegraphics[width=0.7\textwidth]{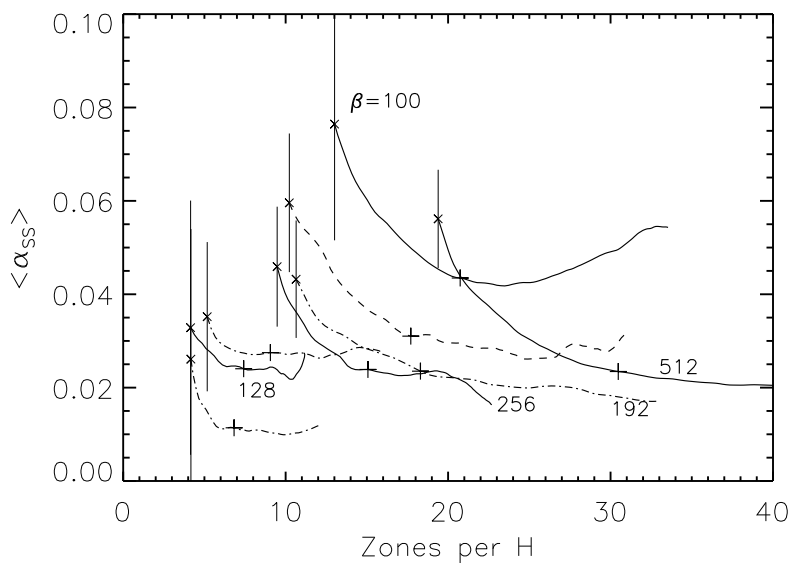}
\end{center}
\caption[]{Time-averaged $\alpha_{SS}$ for the one-loop runs (solid
lines) and the toroidal field runs Toroidal-192, -128 and -64(dot-dashed lines),
as a function of zones per $H$.  The toroidal field runs are
time-averaged from $t=10^4$--$1.4\times 10^4 M$.  The fiducial
two-loop run is include for comparison (dashed line).  The point
marked by a cross on each curve indicates the value of $\alpha_{SS}$
at $R=10M$.
}
\label{fig:alphass-one}
\end{figure}

In the turbulent evolution phase, the one-loop models all have
stress levels that are below the lowest resolution two-loop model.
The reduced stress in the one-loop models also leads to reduced heating
and smaller disk scale heights relative to the two-loop models.
Figure~\ref{fig:alphass-one} plots the time averaged $\alpha_{SS}$
as a function of zones per $H$ for the one-loop models.  Although the
stress is reduced compared to the two-loop models, within the body of
the disk $\alpha_{SS}$ is similar for all three
resolutions.  For example, at $R=10M$, $\alpha_{SS} = 0.024$, 0.024,
and 0.023 for the 128, 256 and 512 resolution models, even though the
zones per $H$ vary from 7 to 30 at that radius.  Resolution has a more
obvious impact near the ISCO, where the number of zones per $H$ is 4,
9, and 19, and $\alpha_{SS} = 0.032$, 0.046 and 0.056.  If we compute
$N_z$ using equation (\ref{eqn:qz}) and the time-averaged values from
the simulations, we find that adequate resolution would require 70, 37,
27 vertical zones at the ISCO.  The largest number of zones would be
required for the Oneloop-128 run because its magnetization 
is lower compared to the other simulations. All three are adequately resolved
by the $\Qp$ criterion.  At $R=20 M$, $H$ is larger and is spanned by 50,
22 and 11 zones, with the $N_z$ criterion calling for 29, 39, and 34
zones per $H$.  By both the $\Qz$ and $\Qp$ criteria
the 128 and 256 models are under-resolved,
and the 512 model is only marginally resolved.

\begin{figure}
\leavevmode
\begin{center}
\includegraphics[width=0.7\textwidth]{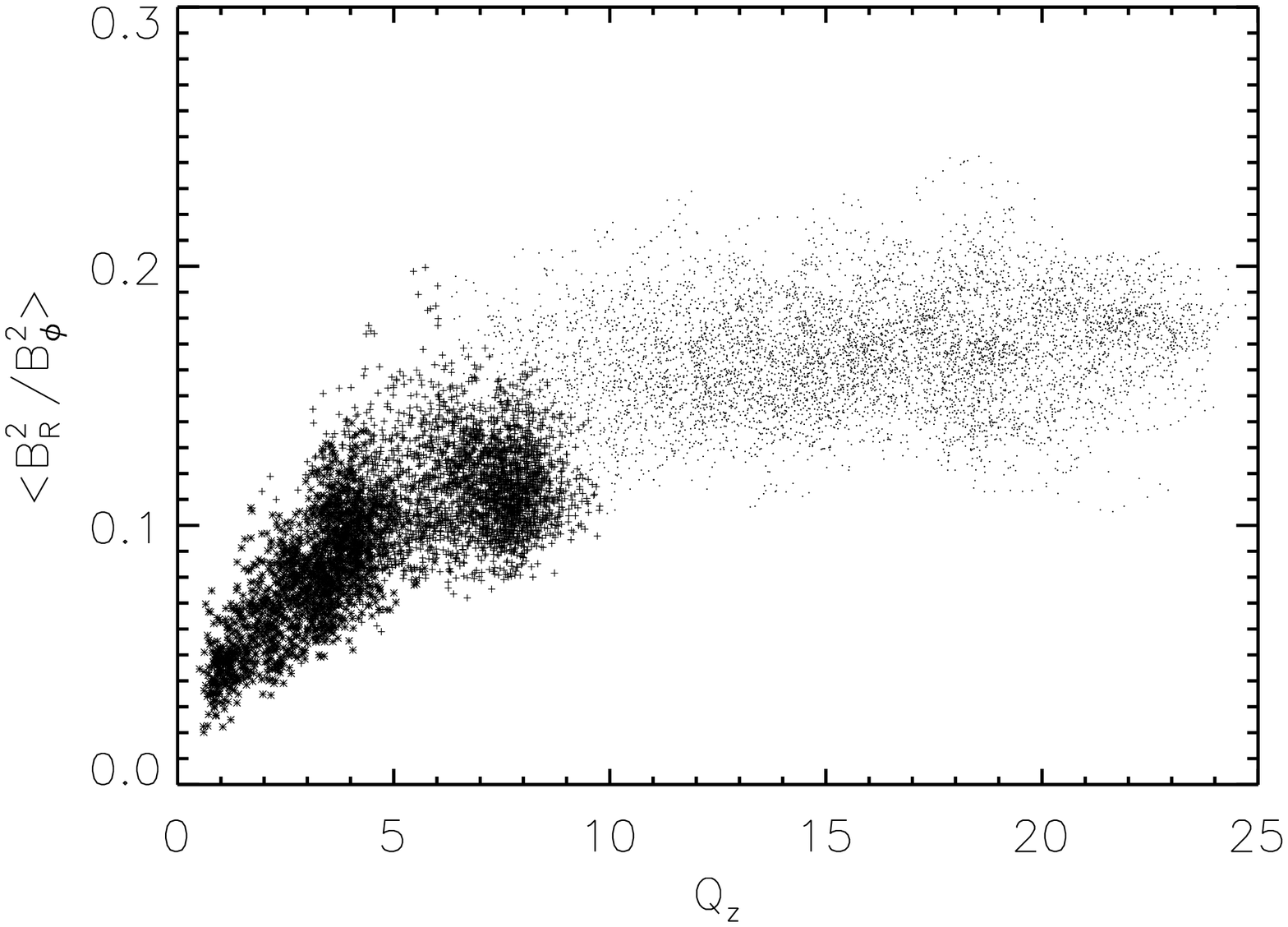}
\end{center}
\caption[]{
Values of $\brbp$ versus $\Qz$ for all radii between $6$--$20M$ and
for times from $t=4000$ to $6150M$.
Stars correspond to the Oneloop-128 run, plus signs to the Oneloop-256
run, and points to the Oneloop-512 run. }
\label{fig:brvsqz2}
\end{figure}

Figure~\ref{fig:brvsqz2} is a plot of $\brbp$ versus $\Qz$ for all radii
between $R=6$--$20M$ and times $t=4000$--$6150M$.  While it has the same
appearance as Fig.~\ref{fig:brvsqz}, as does a time history of the average
value of $\brbp$, Oneloop-512 has a mean $\Qz$ of $16$
and mean $\brbp$ of $0.17$, both lower than Twoloop-512 for the same
time- and radius-interval.

The change in field topology from two- to one-loop, while maintaining the
same average $\beta$, resulted in a reduction of magnetic field energy at
a given resolution.  The total magnetization in the evolved state is
not due to the topology so much as it is due to the initial field strength.
Although the one-loop and two-loop models have equivalent initial
average $\beta$, the different topologies give different initial
$\Qz$; the one-loop configuration has a weaker $B_z$ field.

To explore this further we carry out Oneloop-320b, a simulation with an
average initial field strength of $\beta=100$ and a grid with $320\times
120\times 256$ zones.  This grid maintains $\Delta z$ as in the fiducial
run, sets $\Delta R = \Delta z$, and reduces $\Delta \phi$ to 0.0131.
By increasing the strength of the field, the initial 
$\Qz$ is increased to $\sim 7$, making the linear vertical field
modes adequately resolved.  (This is the strength of the field that was
used in the \cite{Hawley:2001} simulation.)  The evolution proceeds more
rapidly than the $\beta =1000$ models.  The mass on the grid also drops
rapidly; by $t=6662M$ only 10.7\% of the original torus mass remains.
The value of $\alpha_{SS}$ at $R=10M$ is 0.048, twice that
for the $\beta=1000$ one-loop models.
Because of this more rapid evolution, the averaging period for the data
in Table~\ref{table:results} is $t=2000$--$4000M$.

The one-loop configuration lends itself to the production of toroidal field
through shear acting on the initial radial field.  Since Oneloop-320b
has $\beta =100$, the radial field is $10^{1/2}$ times larger than in the
other one-loop models.  This results in a toroidal field with $\beta <
1$ away from the equatorial plane.  In that sense, Oneloop-320b 
is a strong toroidal field model, producing some of the
largest $\Qp$ values in all the simulations.  A similar statement can be
made for the disk thickness, as $H/R$ is second to only the strongest
field two-loop model.  The values of $\Qz$, $\alpha_{mag}$ and $\brbp$
are comparable to the Oneloop-512 model, indicating again that what
constitutes ``adequate resolution'' depends on the initial conditions.
The stress is considerably larger than in all the other one-loop
simulations, as seen in $\alpha_{SS}$ (Fig.~\ref{fig:alphass-one}).

To summarize these results, for one-loop configurations as well as for
two-loop, the best indicators of long-term success in a simulation are
its initial values of $\Qz$ and $\Qp$.

\subsection{Toroidal Field}

We next turn to the evolution of initial toroidal field configurations.
A uniform toroidal field is often used in shearing box initial conditions,
and is of general interest because toroidal fields dominate in an
accretion disk due to the orbital shear.  For example, it is likely
that the $\beta = 1000$ one-loop models are controlled primarily by the
toroidal field generated above and below the equatorial plane.  In the
earliest shearing box simulations \citep{Hawley:1995,Hawley:1995b},
models with net toroidal fields behaved much like those initialized
without a net field.  More recently, \cite{Sorathia:2012} made the same
observation for the cylindrical disk.

A purely toroidal weak field in a differentially rotating plasma is MRI
unstable, although technically it is a transient amplification; $k_r$
the radial wavenumber of a given mode, evolves due to the background
shear \citep{Balbus:1992}.  For a toroidal field, the unstable modes
are nonaxisymmetric, and the most unstable has an azimuthal wavelength
$\sim\lambda_{MRI}$.  The greatest amplification occurs for large $k_z$,
which suggests that high resolution might well be required in both the
azimuthal and vertical dimensions to capture the behavior of the MRI.

Using the same domain as before, including the limited $pi/2$
extent in $\phi$, we consider three initial toroidal
field simulations at different resolutions: Toroidal-64 with  $128\times
64 \times 128$ resolution, Toroidal-128 with $128\times 128\times 128$
resolution, and Toroidal-192 with $256\times 128\times 192$ resolution.
The MRI wavelength, $\lambda_{MRI}$, is the distance an Alfv\'en wave
travels in one orbit, and its size compared to the size of the orbit
is $\lambda_{MRI}/2\pi R = v_A/R\Omega$, the ratio of the Alfv\'en to
orbital speed.  Since the disk is highly supersonic, and the grid zone
size is $R\Delta\phi$, a relatively
strong initial magnetic field is required if $\Qp$ is to be large
enough in the main body of the disk.  
Here we use a uniform toroidal field with $\beta = 10$, defined in
terms of average gas pressure to average magnetic pressure.  Because the
Alfv\'en speed varies in the torus, the initial $\Qp$ also varies
from $\sim 7$--$13$ from the center to the outer radial edge of
the torus using the 64-zone grid; $\Qp$ is doubled for the 128-zone
grid.  We also consider a model with a $\beta
= 100$ initial field strength (Toroidal-128w), for which $\Qp = 5$--7
initially for the 128-zone $\phi$ grid.  And finally, we ran a $\beta=10$
model using second-order interpolation (Toroidal-128s).

\begin{figure}
\leavevmode 
\begin{center}
\includegraphics[width=0.7\textwidth]{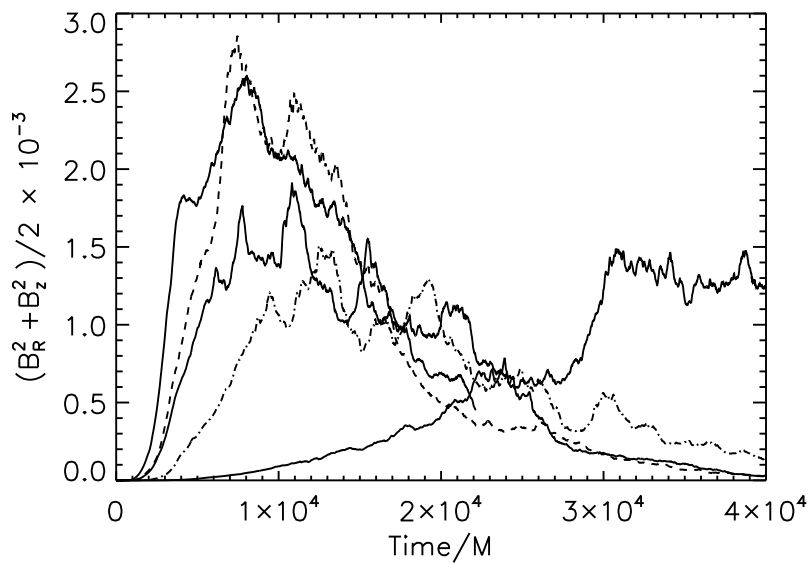}
\end{center} 
\caption[]{Time-evolution of the total poloidal magnetic field energy
for the initial toroidal field models.  From left to right the first
sold curve is Toroidal-192, the dashed curve is Toroidal-128, the next
solid curve is Toroidal-64, the dot-dashed line is Toroidal-128s, which
uses second-order interpolation, and the last curve is Toroidal-128w,
which began with a $\beta=100$ strength field and only achieves a
turbulent state for $t > 3.0\times 10^4 M$.
}
\label{fig:polfield}
\end{figure}

Toroidal field initial conditions have a longer initial MRI growth phase
before fully developed turbulence sets in.  
This is obvious in a plot of the time-history of the poloidal field energy
(Fig.~\ref{fig:polfield}).  The growth rates for the poloidal field
energy (which begins at zero) are highly dependent on the details
of the simulation.  Improved $R$ and $z$ resolution (Toroidal-192)
increases the initial growth rate of the poloidal field components,
but does not significantly alter the peak total energies in the turbulent
state compared to Toroidal-128. Cutting the azimuthal resolution in half
(Toroidal-64) decreases both the growth rate and the peak energy. 

Because of the delay in the onset of turbulence, the diagnostic values for
the $\beta=10$ experiments (Table~\ref{table:results}) are time-averaged
from $t=1.0$--$1.4\times 10^4 M$. In the case of the $\beta=100$ initial
toroidal field model, turbulence developed at a very late time, and the
time-average is taken from $t=3.0$--$4.0\times 10^4 M$.  The quality factor
$\Qz$ is quite generally smaller for the toroidal field models than for
those that begin with a poloidal field.   The $\Qp$ values, on the other
hand, are comparable or larger.  The resolution in $R$ and $z$ seems to
matter less than that of the $\phi$ direction.  When accounting for the
change in $\Delta z$, the $\Qz$ values for Toroidal-192 and Toroidal-128
are comparable, but the reduction to only 64 azimuthal zone reduces $\Qz$
well below the minimum of $\sim 6$.  Similarly $\Qp$ is lower by more than
a factor of two in Toroidal-64.  Only Toroidal-192 has an $\alpha_{mag}$
in the range associated with adequate resolution.  The same can be said
for $\brbp$, except that even for Toroidal-192 the maximum value
is below 0.2 associated with the best-resolved models.  Plotting $\brbp$
versus $\Qz$ as was done in Fig.~\ref{fig:brvsqz} (not shown) again finds
a correlation between the two.  For Toroidal-192 
$\brbp$ flattens out at $\sim 0.15$ as a function of $\Qz$;
$\Qz$ does not exceed $\sim 15$. 

How do these results translate into effective stress?
Figure~\ref{fig:alphass-one} shows time-averaged $\alpha_{SS}$
for the three $\beta=10$ toroidal field runs, time-averaged from
$t=1.0$--$1.4\times 10^4 M$.  The curves for Toroidal-192 and
Toroidal-128 are similar to the $\beta = 1000$ one-loop models, with
$\alpha_{SS} = 0.024$ and $0.027$ respectively at $R=10M$.  Toroidal-64
has a significantly lower value of $\alpha_{SS} = 0.011$.  

The difference between Toroidal-128 and Toroidal-64 indicates that (not
surprisingly) azimuthal resolution is critical.  This is re-enforced
by the experiment where the interpolation scheme is reduced to
second-order (Toroidal-128s).  Unlike the case with the (better-resolved)
two-loop configuration, where the use of second-order had relatively
little apparent effect, here it produces a significant reduction in
growth rate and peak energy, even compared to Toroidal-64 which has
half the azimuthal zones.  The orbital velocity is aligned with the
initial magnetic field, and nonaxisymmetric perturbations experience
numerical diffusion that works against the MRI.  The resulting values
for $\alpha_{mag}$, $\brbp$, and $Q_{z,\phi}$ are comparable to those of Toroidal-64
(Table~\ref{table:results}).  The use of second-order interpolation
reduces the initial MRI growth rates, and the subsequent turbulent state
is affected.  Twoloop-256s appeared unaffected by second-order
interpolation, but it was twice as well resolved in $R$ and $z$ as
Toroidal-128s.  Well-developed MRI-driven turbulence relies on robust
generation of poloidal field, and the increase of diffusion error in
the toroidal direction is compounded by the low poloidal resolution.
The resulting value of $\brbp$ demonstrates this.

Toroidal-128w has an initial field strength of $\beta=100$ and, as such,
has a significantly reduced initial $\Qp$.  Figure~\ref{fig:polfield}
shows that while there is still field amplification, it is at a greatly
reduced rate.  Wavelengths longer than $\lambda_{MRI}$ are still unstable,
i.e., those with low azimuthal wavenumber $m$, but at growth rates that
are reduced $\propto \lambda^{-1}$ compared to the maximum rate.  During
this initial phase, the MRI grows fastest at the edges of the
torus, where the density is low and the Alfv\'en speed is larger compared
to the interior.  It takes until $t=3\times 10^4 M$ for significant
field growth to occur in the core of the disk.  Accretion begins
only beyond this point in time.  The total poloidal field energy at late
time is comparable to that seen in Toroidal-64 (Fig.~\ref{fig:polfield})
near its maximum.  The data in Table~\ref{table:results}, time-averaged
within this period,  reflect a level of magnetization that is consistent
with the Toroidal-64 model.  A reduction in initial field strength is
comparable to a reduction in resolution.

\begin{figure}
\leavevmode 
\begin{center}
\includegraphics[width=0.7\textwidth]{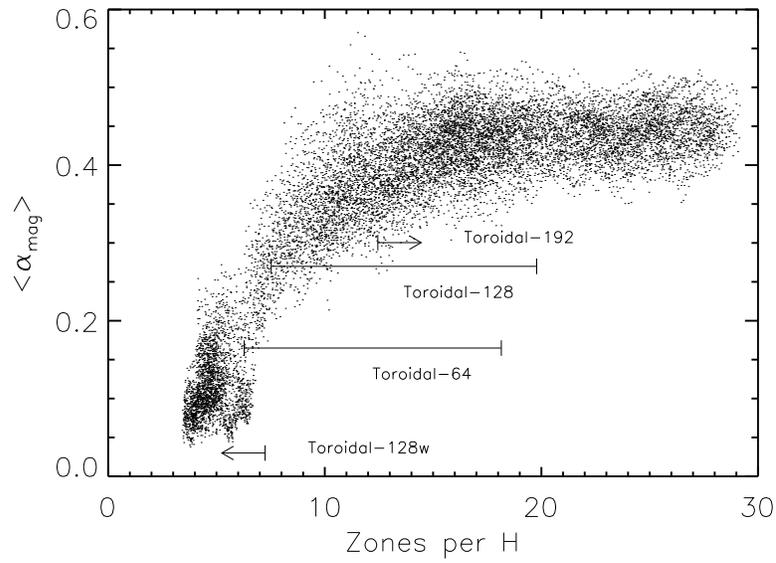}
\end{center} 
\caption[]{
Values of $\alpha_{mag}$ versus number of grid zones per $H$ 
for all radii between $10$--$20M$ and for times from $t=1$--$2\times
10^4M$ for the initial toroidal field runs.
The lines show the range of scale height and the minimum
$\alpha_{mag}$ values for the specific run as labeled.  
The scatter shows the range of variation in $\alpha_{mag}$.  A systematic
decline is observed for resolutions less than 20 zones per $H$.
}
\label{fig:amagtor}
\end{figure}

Figure~\ref{fig:amagtor} examines the behavior of $\alpha_{mag}$ in
these simulations as a function of zones $\Delta z$ per scale height.
The data are individual density-weighted averages at all radii between
$R=10$--$20M$, and all times between $t=10^4$ and $2\times 10^4 M$.
The relationship is striking:  $\alpha_{mag}$ is
reasonably constant for (roughly) 15 zones per $H$ and drops off, at a
very rapid rate for zone number less than 10.  The labeled horizontal
lines indicate the minimum and maximum zone numbers and minimum
$\alpha_{mag}$ for the simulation.  The time interval corresponds to
when the Toroidal-128w run has not yet achieved turbulence throughout
its interior, and it has the lowest $\alpha_{mag}$ values as well as
the smallest number of zones per $H$.

\begin{figure}
\leavevmode 
\begin{center}
\includegraphics[width=0.7\textwidth]{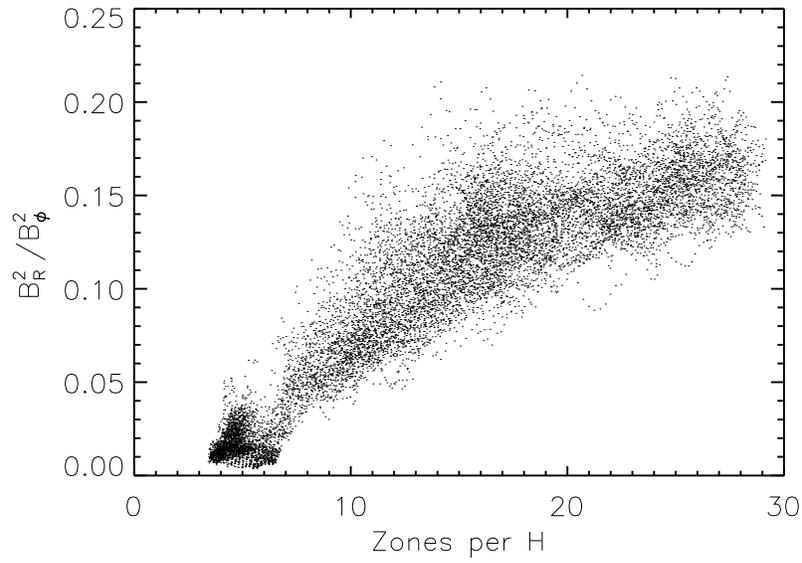}
\end{center} 
\caption[]{
Values of $\brbp$ versus number of grid zones per $H$ 
for all radii between $10$--$20M$ and for times from $t=1$--$2\times
10^4M$ for the initial toroidal field runs.
The scatter shows the range of variation in $\brbp$ which
systematically increases as zones per $H$ increase.
}
\label{fig:brbptor}
\end{figure}

Figure~\ref{fig:brbptor} plots $\brbp$ against the number of
zones per $H$ in the toroidal field runs, over the same radial span
and the same time period.  This shows an even stronger dependence on
resolution than $\alpha_{mag}$.  $\brbp$ remains below 
0.2, but is still increasing as the resolution
approaches 30 zones per $H$.

One notable feature of initial toroidal field global simulations
\citep{Beckwith:2011, Flock:2011} and shearing box simulations
\citep{Simon:2011} is the rapidity with which the net toroidal field
disappears.  For example, Figure 16 of \cite{Simon:2011} shows the
time-evolution of the average toroidal field in shearing box simulations
compared to a similar calculation for a subdomain from a global simulation
of \cite{Beckwith:2011}.  Within approximately 10 orbits the initial net
field is lost and is replaced by a net field of the opposite polarity
and of comparable amplitude to the original net field.  The process then
reverses in an apparent cycle.  This behavior is seen in shearing box
space-time diagrams in $(t,z)$ for field integrated over $(x,y)$, e.g.,
Figure 15 in \cite{Simon:2012}.  \cite{Guan:2011} and \cite{Simon:2011}
examined the time evolution of the mean field in shearing box simulations
and found that it was well described by a generalized $\alpha$-$\Omega$
dynamo model.  The dynamo seen in local simulations also occurs within
the global disk.  \cite{Oneill:2011} also report seeing dynamo action in
their global simulations, noting that the sign of the net toroidal field
alternates on a time comparable to 10 local orbits.  We also see this
behavior in these simulations.  In Toroidal-192, for example, we examined
the time-evolution of $B_\phi$ integrated over a subdomain within the
initial torus centered on $R=15M$, extending from $R=14$--$16M$, and
for $|z| < 5M$.  The initial net field reverses within this subdomain
within 10 orbits.  While there is some indication of periodicity, it is
not as regular as seen in the shearing box. This is not surprising as
the shearing box is, in contrast to a global disk, characterized by a
single frequency $\Omega$.

In sum, much the same resolution criteria hold for simulations with initially
toroidal magnetic field as for those with initially poloidal field.  If the
simulation is to result in fully-developed MHD turbulence, the initial
(and continuing) quality factors must be large enough: $\Qz \gtrsim 15$
and $\Qp \gtrsim 20$.  These can be achieved by different combinations of
initial $\beta$ and grid resolution, but these criteria must be met
one way or another.

\subsection{Transients versus Turbulence}

The simulations presented throughout this paper compare the outcomes using
different resolutions and numerical algorithms, but all runs within a
comparison group begin from a specific idealized, non-turbulent initial
condition.  As a consequence they all experience a linear growth phase
for the MRI followed by the onset of turbulence and the development
of quasi-steady state accretion that lasts until the matter supplied
by the initial condition is depleted.  The properties of the resulting
turbulence might depend on the way the initial linear growth period, with
all its transient effects, depends on resolution.  To what extent are the
reduced turbulent and magnetic energies seen at low resolution due to how
the simulation emerges from the start-up phase, and to what extent is it
due to a given resolution's inadequacy in resolving the turbulent phase
itself?  

Of course, most simulations start from an idealized, and relatively
simple, initial condition, with all the associated initial transient
effects, so this study is representative of the typical case.  But,
it is possible to investigate more directly the influence of resolution
on the turbulent state itself by performing an experiment in which the
quasi-steady state disk is the initial condition.  We start with run a
high resolution, $512\times 128\times 512$, simulation that begins with
the fiducial two-loop initial condition.  This is labeled ``Twoloop-512t''
(for turbulent)  and is evolved well into the quasi-steady state phase.
At time $t=4600M$  the data from this simulation are copied onto
successively coarser grids of $256\times 64\times 256$ (Twoloop-256t)
and $128\times 32\times 128$ (Twoloop-128t).  All three resolutions are
then evolved forward and compared.

To re-grid the data to lower resolution, we use an {\it Athena} restart
file corresponding to time $t=4600M$.  The fundamental data in an
{\it Athena} zone are volume-averaged values of density, momentum and
total energy, and face-centered magnetic fluxes.  The volume element
for a grid zone is $RdRdzd\phi$.  To go from a fine grid to a coarser
grid we do a volume-weighted average over the zones that make up the
single larger zone on the coarser grid.  This conserves total mass and
momentum in the re-grid process.  The magnetic field, on the other hand,
is area-averaged over those zone faces that make up the new larger
zone face.  This preserves net flux and the divergence-free property
of the magnetic field on the new grid.  The regrid process results
in some loss of magnetic energy as oppositely directed fields undergo
numerical reconnection.  The total loss in magnetic energy is 7.2\% in
Twoloop-256t, and 23\% in Twoloop-128t.  Thus, re-griding
introduces by itself a change that will produce a transient response in
the simulation.  Similarly, some kinetic energy is lost by the volume
average onto the new grid.  Kinetic energy is dominated by the orbital
component, which barely changes, but the poloidal kinetic energy decreases
by 2\% and 8\% respectively.  Because of the loss of magnetic and kinetic
energy, we choose to regrid the internal energy and reconstruct the new
total energy on the coarse grid.  This avoids an abrupt change in disk
temperature due to thermalization of the interpolation losses as the
simulation is restarted.

\begin{figure}
\leavevmode
\begin{center}
\includegraphics[width=0.7\textwidth]{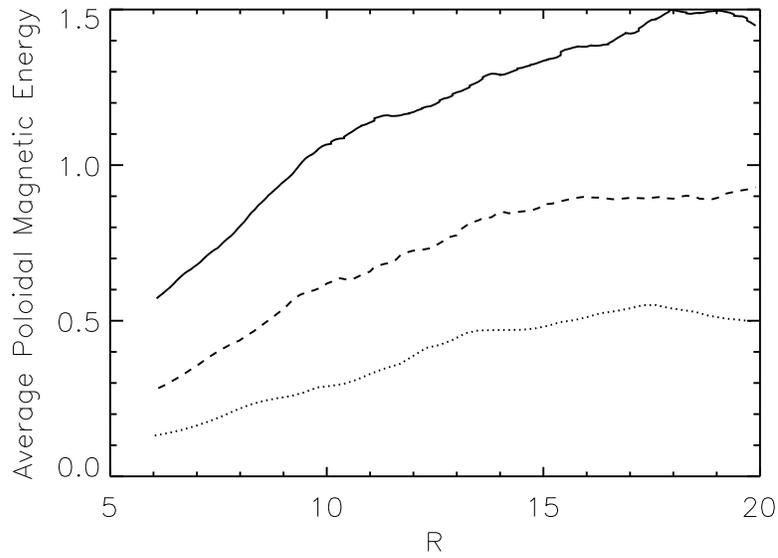}
\end{center}
\caption[]{
The vertically integrated poloidal magnetic field energy ($\times
10^4$) as a function
of radius between $6$--$20M$ time-averaged from $t=5040$ to
$6640M$ for the turbulent restart test runs.
The solid line is Twoloop-512t, the dashed line Twoloop-256t, and the
dotted line is
Twoloop-128t.
}
\label{fig:bpoloidal}
\end{figure}

After restart, the coarse grid simulations relax to a new quasi-steady
state with reduced magnetic energy.  This is rapid, occurring within only
$50M$, showing that it is a local phenomenon as numerical dissipation
occurs at the new, coarse grid resolution.  We consider the interval
between $t=5040$--$6030M$, a point in time after the re-gridded runs
have re-established a quasi-steady state.  Figure~\ref{fig:bpoloidal}
shows the time-averaged vertically integrated poloidal magnetic field
energy as a function of radius.  The coarser grids simply cannot
sustain as high a magnetization, a conclusion that follows without
regard to the evolution history.

This is reflected in the diagnostic values, which are listed in
Table~\ref{table:results}. Again these are time-averaged between
$t=5040$--$6030M$.  Comparison of these diagnostic values with the values
in similarly resolved grids run from the initial state (Twoloop-256 and
Twoloop-128) show great similarity. Again, because the simulations relax
to this state on a relatively short time, it seems that the observed 
level of turbulence is simply what a given resolution can support.

\begin{figure}
\leavevmode
\begin{center}
\includegraphics[width=0.7\textwidth]{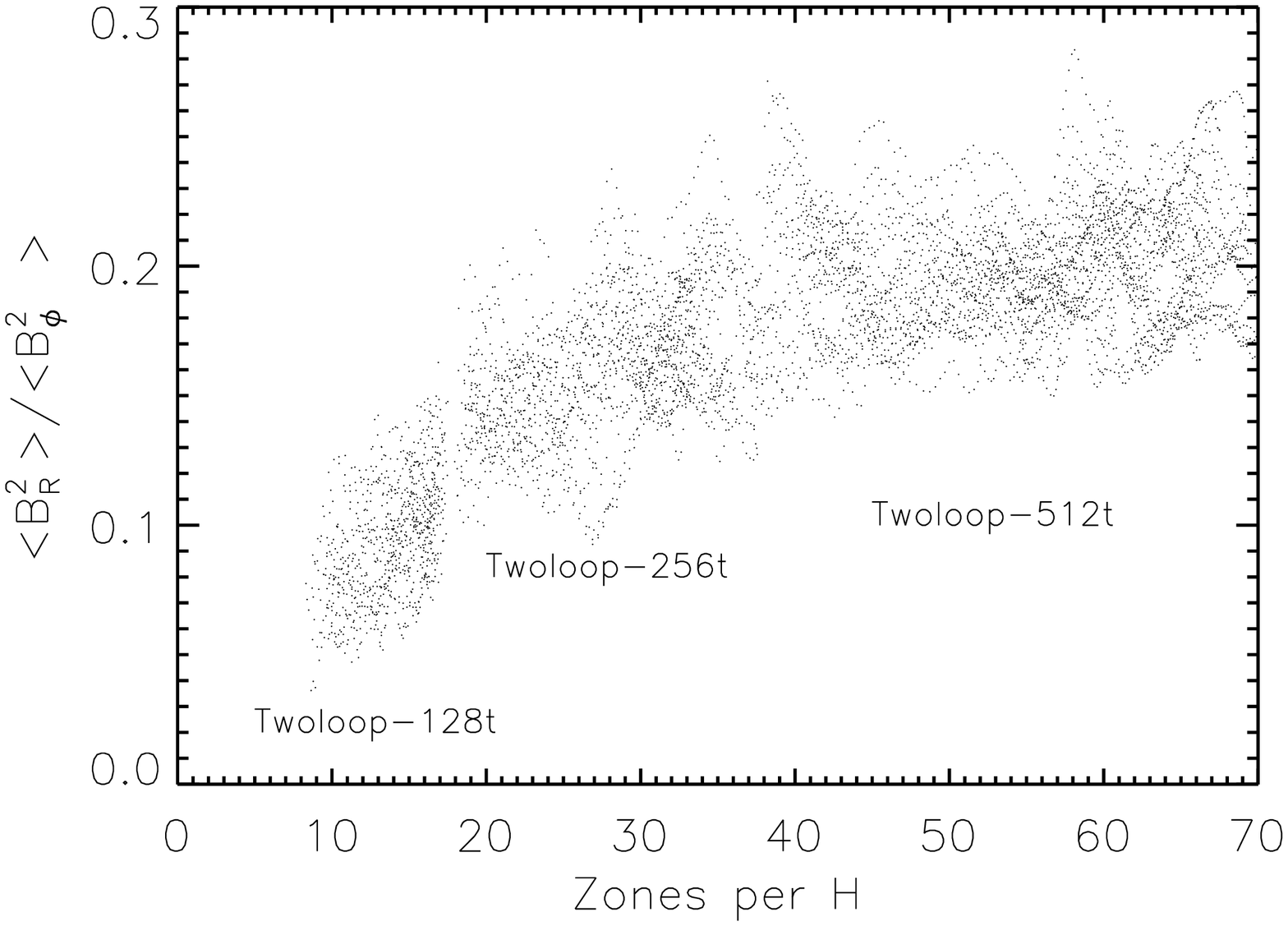}
\end{center}
\caption[]{
Values of $\brbp$ versus number of grid zones per $H$ 
for all radii between $10$--$20M$ and for times from $t=5040$ to
$6640M$ for the turbulent restart test runs.
The three different runs inhabit distinct regions of the plot, labeled
Twoloop-128t, Twoloop-256t and Twoloop-512t. The scatter shows the
relatively wide range of variation in $\brbp$. 
}
\label{fig:brbpvsntest}
\end{figure}

Figure~\ref{fig:brbpvsntest} plots the value of $\brbp$ versus zones per
scale height $H$ for these three simulations.  It should be compared
with Fig.~\ref{fig:brbpvsn}.  The distribution of points is similar
in the two plots, although here the disks are modestly thicker and
the points are shifted to higher numbers of zones per $H$ compared
to Fig.~\ref{fig:brbpvsn}.  This is a difference from the equivalently
resolved models that were run from the initial condition: the thicker disk
should give make up somewhat for low resolution.  However, the overall
behavior of the $\brbp$ diagnostic with effective resolution remains
the same.  Small scales are important in fully developed MRI-driven
turbulence, and the quality parameters, $Q$, provide a measure of a
grid's ability to resolve those features adequately.

\section{Conclusions}

One of the most important scientific goals of global accretion disk
simulations is the ability to make detailed observational predictions
from first-principles physics.  Although the thermodynamics employed
in global simulations remains primitive and other important dynamical
elements like radiation forces have yet to be implemented outside
shearing-box models, in recent years several groups have initiated
programs to use the bolometric energy budgets of simulations for
this purpose \citep{Noble:2011, Kulkarni:2011, Zhu:2012, Schnittman:2012}
The quantitative quality of these predictions depends, of course, on
the quality of the simulation data underlying them.  Any parameter
inferences (e.g., the black hole spin parameter) will be subject to
a systematic error traceable to deficiencies in the resolution data.

In this paper, we have attempted to clarify the computational and numerical
limitations associated with global simulations at the limits of resolution
achievable with contemporary computational resources.  We have found that
resolution and numerical algorithm, as
well as initial field strength and topology, can affect the quantitative
values of MRI-driven accretion in global simulations.  Considerable care
is needed to sort through these effects when interpreting simulation
outcomes.

In particular, the quality factors $\Qz$ and $\Qp$ are valuable measures
of resolution, for both the initial conditions and in the fully developed
MRI-driven turbulence.  Low $Q$ values are unambiguous markers of
inadequate resolution and, consequently, uncertain quantitative values.
Low resolution leads to reduced stress, incompletely developed turbulence,
and reduced accretion rates, but not necessarily a complete cessation
of accretion.  However, when the resolution is especially coarse or there
is an especially large downward fluctuation in the magnetic intensity,
the turbulence may die altogether.  \cite{Shiokawa:2012} carried out a
convergence study of a general relativistic accretion simulation and
observed just this behavior in their lowest resolution simulation
consisting of $96\times 96\times 64$ zones in $(r,\theta,\phi)$.  

By contrast, high $Q$ values indicate adequate resolution.  Moreover,
because adequate resolution leads to larger magnetic saturation levels,
sufficiently high $Q$ values in the initial state can lead to still
higher values as the simulation proceeds.  The key question we have
attempted to answer is, of course, how to define the threshold of
``adequacy", beyond which there are only insignificant changes in the
simulation.  \cite{Hawley:2011} estimated minimum values of $\Qz \ge 10$
and $\Qp \ge 20$.  These values were incorporated into (\ref{eqn:qz})
and (\ref{eqn:qy}) to estimate the required number of grid zones for
a simulation.  In this paper we found that these equations provide
a reasonable estimate of the number of zones needed for adequate
resolution.  A modest increase in the target $\Qz$ and $\Qp$ would
probably provide a better estimate.  Figure~\ref{fig:brvsqz} suggests
$\Qz \ge 15$, for example.  This is broadly consistent with the results of
\cite{Shiokawa:2012}, who argued that their highest resolution simulation,
at $384\times 384\times 256$ zones, was nearly adequately resolved; it
featured $Q$ values equivalent to $\Qp \sim 22$ and $\Qz \sim 9$--29 in
the near hole region.

The threshold values of the two quality factors are not independent
of each other; higher $\Qp$ values can compensate for low $\Qz$.
\cite{Sorathia:2012} suggest that $\Qz > 10$--15 is required for $\Qp
\approx 10$, and that $\Qp > 25$ compensates for low $\Qz$.  The latter
conjecture is supported by our toroidal field runs where $\Qz$ was
clearly low, but with $\Qp \sim 30$ turbulence was nevertheless sustained
with good $\alpha_{mag}$ and $\brbp$ values.  We did not have an example
with a large $\Qz$ and a low $\Qp$; we did not perform any simulations
with a net vertical field and a low $\phi$ grid count, the most likely
scenario where that could be observed.

The present simulations also provided considerable data with which to
test the behaviors of two other simulation quality indicators describing the
physical character of the turbulence rather than the resolution, $\alpha_{mag}$
(which has the equivalent information as the tilt angle used in other
studies), and the ratio of radial to toroidal magnetic field energy,
$\brbp$.   Although these two diagnostics are related, they are not equal.
The tilt-angle, or $\alpha_{mag}$, is the ``efficiency'' of a field
configuration by measuring the degree of correlation between $B_r$
and $B_\phi$, whereas $\brbp$ measures the ability of the turbulence to
generate poloidal field, through the MRI-dynamo process.  We find
that $\alpha_{mag} < 0.4 $ indicates inadequate resolution.  There is a
relation between $\alpha_{mag}$ and the number of grid zones per vertical
scale height; above 20 zones per $H$, $\alpha_{mag}$ levels off at around
0.45 which corresponds to a magnetic tilt angle of $13^\circ$.  $\brbp$
also shows a dependence with resolution, but one that is more demanding
than seen with $\alpha_{mag}$.  $\brbp$ levels off at $\approx 0.2$
in the best resolved case when $H/\Delta z \ge 35$.
This is consistent with the consensus emerging from local simulations
that adequate resolution requires greater than 32
zones per $H$.  Finally we note that both $\brbp$ and $\alpha_{mag}$
are proportional to $\Qz$ until they reach their limiting values.

Based on these criteria, of the simulations we have done for this
paper, excluding the numerical algorithm tests, only the Twoloop-512,
Twoloop-512t,
and Twoloop-256b simulations are adequately resolved, and the later
is resolved by virtue of its relatively strong initial magnetic field.
Twoloop-256, Twoloop-256l, Oneloop-512, Oneloop-320b, and Toroidal-192
are marginally resolved and the other simulations are under-resolved to
varying degrees.  Because our criteria are similar to those of
\cite{Hawley:2011}, we support their judgment of which published
thin disk simulations are close to adequate resolution: only the
set reported by \cite{Noble:2010} meet the test; because the vertical
cell size in the simulations of \cite{Penna:2010} are $\simeq 4\times$
larger, it is likely that they fall short.

In accretion disks the $R$--$\phi$ stress is of fundamental importance,
and stress levels can be affected by resolution.  The time-averaged
$\alpha_{SS}$ values in the adequately resolved $\beta = 1000$ two-loop
simulations are $\alpha_{SS} = 0.030$--0.035, out in the main body of
the disk from $R=10$--$20 M$.  These values are consistent with shearing
box results \citep[e.g.][]{Simon:2012} with the caveat that the $\alpha$
values reported for shearing boxes typically include the Reynolds as
well as the Maxwell stress.  For the under-resolved simulations, however,
$\alpha_{SS} =0.02$--0.029.

Stress can be increased in the models by using stronger
initial fields.  Twoloop-256b has an initial $\beta=250$ and has a
time-averaged $\alpha_{SS} = 0.072$ at $R=10M$; Oneloop-320b, with an
initial $\beta =100$ has $\alpha_{SS} = 0.048$.  An increase in stress
with initial field has also been seen in shearing boxes that have net
fields \citep{Hawley:1995}.

Although our focus in this paper is on the stress in the main part of the
disk, we do observe significant non-zero stress near the ISCO, and these
ISCO stresses are particularly sensitive to resolution.  For example,
the time-averaged $\alpha_{SS}$ at the ISCO are 0.034, 0.065 and 0.076
for the 128, 256 and 512 resolution two-loop $\beta=1000$ simulations
(see Fig.~\ref{fig:alphass}).  Under-resolved simulations
can significantly under-value the near-ISCO stresses compared to
simulations that are adequately resolved in terms of the qualilty
metrics.

In considering the stress levels found in these simulations, one should
bear in mind that we are considering the {\it numerical} convergence for
a particular physical model, here inviscid, ideal MHD without radiation
physics.  Simulations that use a different physical model could get
$\alpha_{SS}$ values that are different from those found here.  Regardless
of the physical model used, however, one should demonstrate that the
resolution used in a given simulation is sufficient so that stress levels
are no longer significantly changed by increasing resolution.

While the number of zones per scaleheight in the disk is an important
metric, the strength of the (initial) magnetic field is also important.
Here $Q$ provides valuable guidance as a better way
to characterize the initial field than some initial averaged $\beta$.
If $Q$ is small initially, the MRI will have difficulty establishing
itself and sustaining a steady state.  Any reduced turbulence in
the initial evolution carries over to the turbulent
evolution of the disk. This effect is amplified where heating occurs.
Stronger MRI turbulence leads to stronger heating, increasing
the scale height $H$, and hence the number of zones per scaleheight.
The simulations that were adequately resolved initially became more so
as a consequence of their evolution.

The effect of resolution on the level of magnetization and turbulence in
the quasi-steady state is intrinsic; reduced turbulence levels
are not solely the result of a failure to properly resolve the initial
linear MRI.  When we take a well-resolved turbulent disk and remap it
onto lower resolution grids, the magnetic energy rapidly readjusts to new,
lower level.  Coarser grids cannot sustain as high a level of
magnetization, regardless of the evolution history.

We have found that adequately resolved global simulations have
$\alpha_{SS}$ values that are consistent with shearing box simulations
that use the same physics.
We also found another important similarity between global and local
models: global simulations demonstrate clear evidence of dynamo behavior
as observed in the shearing box.  This agrees with \cite{Oneill:2011} who
observed dynamo behavior in their global simulations.  The dynamo is less
coherent than seen in shearing boxes, however.  Global simulations lack
any special orbital frequency, unlike the shearing box, and experience
secular evolution with time.

Although MRI-driven turbulence in global simulations is very similar
to turbulence in a shearing box, there are specific effects discernible
only with a global treatment.  One of the most obvious is the change
with radius of disk properties, notably including the consistent sharp
rise in stress toward the ISCO when there is adequate resolution.
As discussed by \cite{Sorathia:2012} global models differ from local
models in a fundamental way due to the secular evolution of the former,
at least for global simulations to date.  The timescale for the
secular evolution of the disk is much shorter than the timescale for the
establishment of steady state MHD turbulence.  The initial evolution in
shearing box simulations lasts at least 10 orbits, and characteristic
values are usually derived from boxes that have been run for times
greater than 50 orbits.  In this sense, not only do global disks tend
to be under-resolved relative to shearing boxes, they also tend to be
under-evolved.  This represents a significant challenge, both in total
grid size and in evolution time.  Even though one can run hundreds of
orbits at the ISCO, the inner disk must be fed from an outer disk that
evolves on a much longer timescale.  A long term steady state in an inner
accretion disk requires that it be fed at a steady rate from large radius
in a self-consistent manner and for a significant number of orbits.

Looking forward, it appears that adequate resolution can be achieved
in global simulations, but it requires a considerable number of zones.
Because computational resources are always finite, they must be placed
adroitly in order to have the most positive impact on the quality
of the simulation.   We recommend:
\begin{itemize}

\item When the disk aspect ratio $H/R$ is constant, spherical coordinates
are best, as they allow one to maintain a constant number of zones per $H$,
independent of $R$.  If instead the disk scale height $H$ is constant,
cylindrical coordinates would be preferable for the same reason.

\item Use of the HLLD flux solver rather than a diffusive solver like HLLE.
The use of a higher-order reconstruction scheme can be important for
improved results, particularly at marginal resolution.

\item The initial conditions of any simulation must be well-resolved
as indicated by $\Qz$ and $\Qp$.

\item Adequate resolution in both the poloidal and azimuthal directions
is essential.  In terms of the quality factors we have defined, this
means $\Qz \gtrsim 10$--15 and $\Qp \gtrsim 20$, although an especially
high value in one of the $Q$ parameters can compensate for a slightly
low value in the other.  These values must be maintained at adequate
levels throughout the simulation; this goal can generally be achieved
when there are at least $\simeq 35$ vertical cells per scale height in
the disk body.  All these criteria apply to all the magnetic topologies
we have examined so far: single poloidal loops, dual poloidal loops,
and purely toroidal initial field.

\end{itemize}

\section*{Acknowledgements}

This work was partially supported by NASA grant NNX09AD14G and NSF
grant AST-0908869 (JFH) and NSF grant AST-0908336 (JHK).  Some of the
simulations described here were carried out on the Kraken system at NICS,
supported by the NSF.  We thank Jake Simon for comments on this work.
We thank the referee for suggesting the regrid tests described in
\S 4.5.

\bibliographystyle{apj}
\bibliography{hawleyreferences}

\end{document}